\documentstyle[eqsecnum,aps,prb,multicol,epsfig]{revtex}

\begin{document}

\title{Velocity-force characteristics of an interface driven through a
  periodic potential}
\author{A.M. Ettouhami$^1$ and Leo Radzihovsky$^{1,2}$}
\address{$^1$ Department of Physics, University of Colorado, Boulder,
  CO 80309}
\address{$^2$ Department of Physics, Harvard University, Cambridge, MA 02138}
\date{\today}
\maketitle

\begin{abstract}
  
  We study creep dynamics of a two-dimensional interface driven
  through a periodic potential using dynamic renormalization group
  methods. We find that the nature of weak-drive transport depends
  qualitatively on whether the temperature $T$ is above or below the
  equilibrium roughening transition temperature $T_c$.  Above $T_c$,
  the velocity-force characteristics is Ohmic, with linear mobility
  exhibiting a jump discontinuity across the transition.  For $T \le
  T_c$, the transport is highly nonlinear, exhibiting an interesting
  crossover in temperature and weak external force $F$. For
  intermediate drive, $F>F_*$, we find near $T_c^{-}$ a power-law
  velocity-force characteristics $v(F)\sim F^\sigma$, with
  $\sigma-1\propto \tilde{t}$, and well-below $T_c$, $v(F)\sim
  e^{-(F_*/F)^{2\tilde{t}} }$, with $\tilde{t}=(1-T/T_c)$. In the
  limit of vanishing drive ($F\ll F_*$) the velocity-force characteristics
  crosses over to $v(F)\sim e^{-(F_0/F)}$, and is controlled by
  soliton nucleation.
\end{abstract}


\begin{multicols}{2}

\section{Introduction}

The problem of elastic media pinned by an external potential provides
a unifying framework for understanding a large number of condensed
matter phenomena, such as for example surface growth
\cite{Krim,Kardar}, nonlinear transport in anisotropic metals
\cite{Gruner}, dissipation in superconductors
\cite{Blatter-et-al,Giamarchi3,BMR,Scheidl-Vinokur}, Wigner crystals
\cite{Andrei}, earthquakes \cite{Fisher_earthquakes} and friction
\cite{Robins}. Randomly-pinned elastic systems are also toy models for
considerably more complicated problems of glasses \cite{BinderYoung}.
Much of the recent interest in such problems has been rekindled by the
discovery \cite{Bednorz-Muller} of high-temperature superconductors
(HTSC), and efforts to understand the nature of their $B-T$ phase
diagram and dissipation controlled by statics and dynamics of elastic
arrays of vortex lines
\cite{FFH,Blatter-et-al,Giamarchi3,BMR,Scheidl-Vinokur}. A combination
of thermal fluctuations, pinning, and external drive leads to a wide
range of new and interesting collective phenomena that is common to
many physical realizations of elastic media.

There has been considerable progress in understanding the static
properties of pinned elastic media
\cite{Blatter-et-al,DSFisher,BraggGlass,Nattermann-Scheidl}. Much of
the recent interest has therefore shifted to dynamics, with current
focus on nonequilibrium, driven dynamics of these rich systems
\cite{Blatter-et-al,Giamarchi3,BMR,Scheidl-Vinokur}. Once the elastic
medium is driven, however, many new questions arise, such as the
governing nonequilibrium equation of motion, phase classification and
stability, nature of the corresponding phase transitions, and the
resulting nonequilibrium phase diagram
\cite{Giamarchi3,BMR,Scheidl-Vinokur,BalentsFisher,RTcdw}.  Among
these many challenging questions, the velocity ($v$)\, --\, force
($F$) characteristics (the IV curve, in the context of superconductors
and charge density waves (CDW's)) is the observable that is most
directly accessible experimentally \cite{Bhattacharya} and is therefore
of considerable theoretical interest.

Despite of considerable richness of many aspects of the driven state
\cite{Giamarchi3,BMR,Scheidl-Vinokur,RTcdw,Bhattacharya,ChenBalentsFisherMarchetti},
at large drives the velocity-force characteristics of a uniformly
sliding medium approaches Ohmic form with deviations $\delta v$ that
can be computed perturbatively \cite{SchmidHauger} in the ratio
$\delta v/v$. At zero temperature, if elastic
\cite{Middleton_uniqueness}, the medium is pinned for drives smaller
than a critical $T=0$ value $F_c$, and undergoes a nonequilibrium
depinning transition to a sliding state, with $v\sim |F-F_c|^\beta$
playing the role analogous to an order parameter distinguishing pinned
and sliding phases \cite{DSFisher_depinning}.  Finite temperature
rounds the depinning transition \cite{Middleton_finiteT}, allowing
activated creep motion of the elastic solid even for drive far below
$F_c$.

Much of the insight about this highly nontrivial creep regime that is
the focus of our work comes from a scaling theory of depinned droplet
nucleation
\cite{FFH,FisherHuse,Ioffe-Vinokur,Feigelman-et-al,Nattermann}.  This
approach generically predicts collectively pinned elastic media to
exhibit a highly nonlinear $v(F)$, with a vanishing linear mobility,
corresponding to transport activated over barriers that diverge with
system size and vanishing drive. Recently, in the case of random
pinning, these scaling predictions have been put on firmer ground
through a detailed dynamic functional renormalization group (DFRG)
calculations of $v(F)$ \cite{LRaps,Chauve_etal}, which
indeed predict $v(F)\sim e^{-1/F^\mu}$, with a universal $\mu$
exponent.  However, in the case of random pinning a number of
technical problems with DFRG remain, precluding a fully controlled
analysis \cite{BFRunpublished,BalentsLedoussal}.

It turns out, however, that many problems of interest, such as surface
growth \cite{Wolf-et-al}, 2D colloidal crystals in periodic potentials
\cite{Clark,Leiderer,FNR,RFN}, and vortices pinned by artificial dot
arrays \cite{Mishalkov_etal} or by intrinsic pinning in e.g., HTSC,
involve the considerably simpler but still nontrivial problem of {\em
  periodic} pinning. In addition to addressing numerous interesting
physical problems, study of motion in a periodic potential provides a
nice laboratory to explore new calculational methods.

A driven sine-Gordon model is the simplest description of such
periodic pinning problem, with an important simplifying feature of
absence of topological defects such as dislocations or phase slips
that can be important for understanding the dynamics of vortex arrays
and CDW's \cite{BMR,MarchettiMiddleton}. Directly applicable to
crystal growth phenomena, this model has been extensively studied in
the literature
\cite{Jose,Chui,Wiegmann,Ohta,Amit,Knops,Neudecker,Nozieres,RostSpohn,vonBeijeren}.
In equilibrium, among many other things, it describes the famous
crystal surface roughening transition from the low-temperature smooth
phase with bounded surface roughness to the high-temperature rough
phase with logarithmic height correlations.

One of the many interesting questions that naturally arises is: {\em
  What are the signatures of the roughening transition in the driven
  transport~?}  More specifically here we are interested in qualitative
differences (if any) in $v(F)$ above and below $T_c$.  The equilibrium
limit of this question was addressed in the classic analysis of the
equilibrium dynamics by Chui and Weeks \cite{Chui} and by Nozi\`eres and
Gallet \cite{Nozieres}. By extending the standard analysis to
equilibrium {\em dynamics} they found the vanishing of the linear
mobility below the roughening transition, consistent with the static
picture of the smooth phase where the interface is pinned by the
periodic potential.

In the presence of an external drive, the sine-Gordon model was studied sometime
ago by Hwa, Kardar, and Paczuski \cite{Hwa} and by Rost and
Spohn \cite{RostSpohn}.  Although these works led to considerable
progress, computing the renormalization group (RG) flow equations 
for model parameters at nonzero $F$, they
concentrated mainly on the influence of Kardar-Parisi-Zhang
(KPZ) \cite{KPZ} nonlinearities (not considered in previous
studies \cite{Nozieres}) important at strong external drive, but said
little about the actual $v(F)$ characteristics of the driven interface
in the $F\to 0$ creep regime.
The driven sine-Gordon model has also been considered by
Blatter et {\em al.} \cite{Blatter-et-al} in a complementary approach
via a high velocity perturbative expansion for $v(F)$. Consistent with
Refs. \onlinecite{Chui} and  \onlinecite{Nozieres}, these last authors 
found that while the correction
$(\delta v/v)$ remained finite above $T_c$, it diverged below $T_c$,
thereby suggesting nontrivial transport changes across the roughening
transition, but leaving the form of $v(F)$ in the creep regime an open
problem. Of course, because at finite temperature the interface can move for
any finite drive $F$, at sufficiently long scales the periodic
potential is averaged away at both low and high temperatures, thereby
leading to the rounding of the roughening transition itself.
Nevertheless, we expect that the velocity-force characteristics in the
{\em creep} regime is controlled by the equilibrium physics and
precise qualitative distinction of $v(F)$ in the rough and smooth
phases should exist \cite{Remark1}.

\begin{center}
\begin{figure}[b]
\includegraphics[scale=0.60]{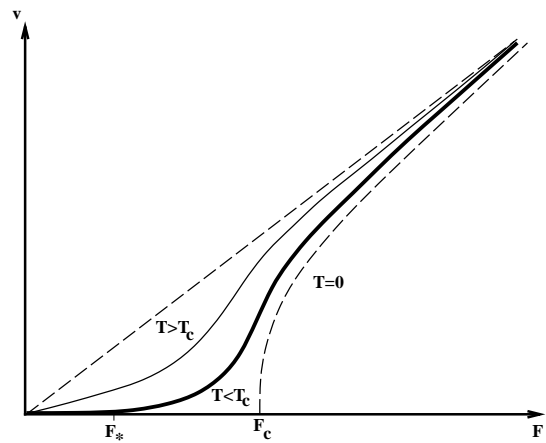}
\caption[Typical velocity-force characteristics of a driven
  interface in a periodic potential]{
  Typical velocity-force characteristics of a driven
  interface in a periodic potential.  At zero temperature, the
  interface remains pinned ($v=0$) until $F$ reaches the critical
  force $F_c=pg$. At finite temperatures, we find that the
  near-equilibrium response of the interface to a small ($F\ll F_c$)
  driving force depends on whether $T$ is above or below the
  roughening temperature $T_c$. For $T>T_c$, the velocity-force
  characteristics is Ohmic ($v(F)\sim F$) down to $F=0$, while for
  $T<T_c$ and forces smaller than a characteristic force $F_*$,
  the characteristics is strongly non-linear, $v\sim
  \exp\big(-F_0/F\big)$, creep motion via activation over barriers
  that diverge in a vanishing drive limit. 
}\label{fig_vfchar}
\end{figure}
\end{center}

In this paper, our goal is to understand in detail the physical
consequences of the divergences found in the high-velocity
perturbative expansion and in particular to compute the creep
velocity-force characteristics in both phases and across the
roughening transition, utilizing dynamic RG
\cite{BMR,LRaps,BFRunpublished,BalentsLedoussal,Chauve_etal}.  Consistent with
perturbative analysis, we find that the nature of transport depends
qualitatively on whether the temperature is above or below the
equilibrium roughening transition temperature $T_c$. Above $T_c$, the
velocity-force characteristics is Ohmic, with the mobility remaining
finite for $T\to T_c^+$. In contrast, for $T<T_c$, we find that the
linear mobility vanishes on long length scales, and therefore exhibits
a nonuniversal jump discontinuity across the roughening transition
\cite{Nozieres,RostSpohn,Blatter-et-al}.  In the smooth phase, the
transport is a strongly nonlinear function of applied force, showing a
rich universal crossover in temperature and applied force (Fig.
\ref{fig_vfchar}).  At an intermediate drive $F>F_*(\tilde{g},T)$, larger
than the pinning ($\tilde{g}$)- and temperature-dependent
strong-coupling crossover force
\begin{equation}
F_*(\tilde{g},T)\sim\left\{\begin{array}{ll} 
    e^{-b_1/\tilde{g}^2},\qquad & T\to T_c^-\\
    \tilde{g}^{1/\tilde{t}},\qquad & T\ll T_c
           \end{array}  \right. 
\label{F*}
\end{equation}
the velocity-force characteristics strongly depends on the level of
proximity to $T_c$, with:
\begin{equation}
v(F)\sim\left\{\begin{array}{ll} F^{(1+b_2\tilde{t} )},\qquad & T\to T_c^-\\
                    e^{-(F_*/F)^{2\tilde{t} }},\qquad & T\ll T_c
                                           \end{array}  \right. 
\label{vF>F*},
\end{equation}
where $\tilde{t}=(1-T/T_c)$, and $b_1$ and $b_2$ are nonuniversal
constants of order unity.  For sufficiently low drive
$F<F_*(\tilde{g},T)$, the motion is instead always via activated
soliton creep, with the velocity-force characteristics crossing over
to 
\begin{equation}
v(F)\sim e^{-F_0/F}\quad,\qquad F < F_* ,
\label{vF<F*}
\end{equation}
with $F_0$ another characteristic force that will be defined below,
Eq.(\ref{F_0}).

This paper is organized as follows. We introduce the driven
sine-Gordon model in Section \ref{model} and analyze it in Section
\ref{secPT} using simple perturbation theory in the
\noindent
pinning potential strength. While for weak pinning this computation is
convergent for $T>T_c$, it fails for arbitrarily weak pinning in the
smooth phase. In section \ref{DynamicRG} we employ dynamic RG
techniques to make sense of these divergences, and in Section
\ref{Analysis} we use these results to compute $v(F)$ through the
roughening transition.  We conclude in Section \ref{sine_G_Conclusion}
with a summary of the results and a discussion of open problems and
future directions.

\section{Driven sine-Gordon model}
\label{model}

In equilibrium a two-dimensional sine-Gordon model of an elastic
interface is described by a Hamiltonian
\begin{eqnarray}
H = \int \! d{\bf r}\,\Big[\frac{1}{2}\,K\,(\nabla h)^2 -
g\,\cos\big(p h({\bf r})\big)\Big],
\label{Hamiltonian}
\end{eqnarray}
where ${\bf r}$ is a two dimensional vector in the $(xy)$ plane,
$h({\bf r})$ is the height of the interface above the $(xy)$ plane
(taken to be along the $z$ direction in the embedding space) at
location ${\bf r}$, $K$ is the interfacial surface tension, $g$ is the
pinning strength and $d=2\pi/p$ is the period of the potential.  In
the context of a crystalline surface, with $H$ characterizing its
equilibrium roughness, the periodic pinning potential softly encodes
lattice periodicity of the bulk crystal, corresponding to the $h\to
h+d$ a symmetry of the surface energy, with $d$ being the crystal
lattice constant perpendicular to the interface.

In the absence of any additional conservation laws, long scale
equilibrium dynamics can be described by a simple, relaxational (model
A) Langevin equation
\begin{eqnarray}
\gamma\,\partial_t h & = & -\frac{\delta H}{\delta h({\bf r},t)} 
+ \zeta({\bf r},t) \; ,
\nonumber\\
& = & K\nabla^2h({\bf r},t) - pg\sin\big(ph({\bf r},t)\big) + \zeta({\bf r},t)\; ,
\label{eom_eq}
\end{eqnarray}
where $\gamma$ is the microscopic friction coefficient, and 
$\zeta({\bf r},t)$ a zero-mean, Gaussian thermal noise describing the
interaction of the system with the surrounding heat bath at
temperature $T$, with
\begin{eqnarray}
\langle\zeta({\bf r},t)\zeta({\bf r}',t') \rangle 
= 2\gamma T\delta({\bf r}-{\bf r}')\delta(t-t'),
\end{eqnarray}
in equilibrium imposed by the fluctuation-dissipation theorem (FDT),
forbidding independent renormalization of $T$.

The dynamic description of an interface driven by an external force
$F$ (in the context of crystal growth proportional to the difference
between the chemical potentials of the solid and vapor phases) is
substantially modified. In addition to the obvious addition of the
driving force $F$ on the right-hand side of Eq.~(\ref{eom_eq}),
nonequilibrium dynamics permits the appearance of nonconservative
forces (those not expressible as derivatives of $H$), the most
important of which is the famous Kardar-Parisi-Zhang \cite{KPZ} (KPZ)
$({\nabla} h)^2$ nonlinearity, allowed by the explicit breaking by the
drive of the $z\to -z$ symmetry. An additional important effect of
driving appears as the renormalization of ``temperature'' $T$,
corresponding to the breakdown of the fluctuation-dissipation theorem,
that is, the renormalization of the friction coefficient $\gamma$ is
independent of that of the variance of the noise $\xi$. Even if these
nonequilibrium effects are not recognized a priori, they appear upon
coarse-graining of equation (\ref{eom_eq}) as soon as the external
drive $F$ is included \cite{Remark1,BMR}.  The resulting nonequilibrium
equation of motion is given by \cite{RostSpohn}
\begin{equation}
\gamma\,\partial_t h  
\!=\!  K\nabla^2h \!+\! \frac{\lambda}{2}\big(\nabla h\big)^2
\!-\! pg\sin\big(ph({\bf r},t)\big) + F + \zeta({\bf r},t) \, .
\label{eom_neq}
\end{equation}

Our goal here is to apply the machinery of the dynamic RG to compute
the velocity-force $v(F)$ characteristics for the above model, focusing
on the nontrivial creep regime of the smooth phase, where naive
perturbative expansion in the pinning potential $g$ fails.

\section{Dynamic Perturbation Theory}
\label{secPT}

It is instructive to first study the velocity-force characteristics
through a simple perturbative expansion in the pinning potential $g$.
Starting from equation (\ref{eom_neq}), it is convenient to shift
$h({\bf r},t)=v_0t+u({\bf r},t)$ with $v_0=F/\gamma$ the unperturbed
($g=\lambda=0$) expression of the velocity. Averaging (\ref{eom_neq})
over thermal fluctuations, and ignoring the KPZ term, we find that the
velocity $v$ of the moving interface is given by
\begin{mathletters}
\begin{eqnarray}
v & = & \langle \partial_t h\rangle \\
& = & \frac{F}{\gamma} - \frac{pg}{\gamma}\,\langle\sin\big(pu({\bf r},t) + pv_0t\big)\rangle \quad,
\label{eom_pt}
\end{eqnarray}
\end{mathletters}
where we used the fact that $\langle\zeta({\bf r},t)\rangle=0$. We now let
\begin{eqnarray}
u({\bf r},t) = u_0({\bf r},t) + u_g({\bf r},t) \; ,
\end{eqnarray}
where 
\begin{eqnarray}
u_0({\bf r},t) = \int d{\bf r}'dt'\; R_0({\bf r}-{\bf r}',t-t')\zeta({\bf r}',t')
\end{eqnarray}
is the thermal (noninteracting) part of the interface displacement,
\begin{eqnarray}
u_g({\bf r},t)  =  pg\int d{\bf r}'dt' &{\,}& R_0({\bf r}-{\bf r}',t-t')
\nonumber\\
& \times &\sin\Big(\frac{pFt'}{\gamma} + pu_0({\bf r}',t')\Big)
\end{eqnarray}
is the correction to $u$ linear in the pinning potential strength $g$,
and $R_0({\bf r}-{\bf r}',t-t')=\delta\langle{u_0}({\bf
  r},t)\rangle/\delta F({\bf r}',t')$ is the response function of the
free interface \cite{Landau-Lifshitz}.  Expanding equation
(\ref{eom_pt}) in $u_g$, and averaging over the thermal noise $\zeta$,
we find (see also Appendix \ref{App_sineG_B})
\begin{eqnarray}
v = \frac{F}{\gamma}  & - & \frac{p^3g^2}{2\gamma}\int d{\bf r'}dt'
\,\mbox{e}^{-\frac{1}{2}p^2 C_0({\bf r}-{\bf r'},t-t')} \times
\nonumber\\
&\times&\sin[\frac{pF}{\gamma}(t-t')]\,
R_0({\bf r}-{\bf r'},t-t') \;, 
\label{vfcharPT}
\end{eqnarray}
where $C_0({\bf r}-{\bf r}',t-t')=\langle[u_0({\bf r},t)-u_0({\bf
  r}',t')]^2\rangle$ is the connected correlation function of a free
interface given by
\begin{eqnarray}
C_0({\bf r},t) \simeq \frac{T}{2\pi K}\;\ln\big[
1 + \Lambda^2\big( r^2 + \frac{Kt}{\gamma}\big)\big] \; .
\label{phi1}
\end{eqnarray}

The above velocity-force characteristics, equation (\ref{vfcharPT}),
is most easily evaluated at zero temperature where $C_0({\bf r},t)=0$.
In this limit, using
\begin{eqnarray}
R_0({\bf r},t) = \frac{\theta(t)}{\gamma}\;\int_{\bf q}^\Lambda 
\mbox{e}^{-\frac{Kq^2t}{\gamma}}\;\mbox{e}^{i{\bf q}\cdot{\bf r}} \; ,
\end{eqnarray}
($\theta(t)$ is Heaviside's unit step function) inside
Eq.(\ref{vfcharPT}) and integrating over the time variable $t'$, we
obtain
\begin{eqnarray}
v = \frac{F}{\gamma} - \frac{p^3g^2}{2\gamma}\int d{\bf r}\int_{\bf q}^\Lambda
\frac{pF}{K^2q^4 + p^2 F^2}\;\mbox{e}^{i{\bf q}\cdot{\bf r}} \; .
\end{eqnarray}
In the above equations and througout the rest of this paper we use a
shorthand notation $\int_{\bf q}$ for $\int d{\bf q}/(2\pi)^2$, and
the superscript $\Lambda=2\pi/a$ is the ultra-violet cutoff set by the
in-plane lattice constant $a$, generically distinct from the period
$d=2\pi/p$ perpendicular to the interface.  Performing the integration
over the space variable ${\bf r}$ in the last equation, and using the
resulting Dirac delta-function $(2\pi)^2\delta({\bf q})$ to complete
the ${\bf q}$ integral, we find a $T=0$, leading order (in pinning
$g$) expression for the $v(F)$ characteristics
\cite{Blatter-et-al,SneddonCrossFisher,SchmidHauger}
\begin{equation}
v = \frac{F}{\gamma}\Big( 1 -
\frac{1}{2}\,\big(\frac{F_c}{F}\big)^2\Big)\quad ,\quad F\gg F_c\ ,
\label{vFPTzeroT}
\end{equation}
where $F_c=pg$ is the zero-temperature critical force, in agreement
with the condition $F_c=\mbox{max}\big|\partial V(h)/\partial h\big|$
of disappearance of metastability ($V(h)=-g\cos(ph)$ is the pinning
potential).  As is clear from this result for $v(F)$, even at $T=0$,
the perturbative corrections are small for sufficiently large applied
force $F$ relative to the pinning force $F_c$ (equivalently, for
sufficiently weak pinning $g$ at fixed $F$). In this fast moving
regime, the metastability is absent and pinning gives only
a small correction to the motion with $v(F)$ deviating only weakly from
the pinning-free Ohmic response $v_0(F)= F/\gamma$.  It is reassuring
to note that, since at $T=0$, only the ${\bf q}=0$ mode contributes to
the $v(F)$, Eq.~(\ref{vFPTzeroT}) agrees with the high-drive limit of
the {\em exact} $T=0$ result \cite{Risken,Scheidl_mobility} for a
single particle driven through a one-dimensional sinusoidal potential
\begin{eqnarray}
v(F) = \frac{F}{\gamma}\;\sqrt{1 - \left(\frac{F_c}{F}\right)^2} 
\quad ,\quad F>F_c\ .
\end{eqnarray}
This suggests that the $v(F)$ characteristics of a driven interface
should also exhibit a square-root cusp with an infinite slope at
$F=F_c$. At $T=0$ the interface is strictly pinned for $F\le F_c$.

In contrast, at any finite temperature the interface moves for
arbitrarily weak force and hence there is no sharp depinning
transition.  The perturbative expression for $v$, Eq.~(\ref{vfcharPT})
can be readily evaluated by using the fluctuation-dissipation
relation
\begin{eqnarray}
\theta(t)\,\partial_t C_0({\bf r},t) = 2 T R_0({\bf r},t)
\label{FDT}
\end{eqnarray}
obeyed by the equilibrium response and correlation functions. Using
this relation to eliminate $R_0({\bf r},t)$ from the {\em rhs} of
Eq.~(\ref{vfcharPT}) and integrating by parts over $t'$ we find
\begin{equation}
v = \frac{F}{\gamma}\Big[
1 - \frac{p^2g^2}{2\gamma T}\int \!\!d{\bf r}\!\!\int_0^\infty\!\! 
dt\; \cos(pFt/\gamma)\mbox{e}^{-\frac{1}{2}p^2 C_0({\bf r},t)}\Big]\, .
\label{vfcharPT2}
\end{equation}
Inserting into this last equation the expression of the correlation
function $C_0({\bf r},t)$ of a harmonic interface given in
Eq.~(\ref{phi1}) leads to
\begin{equation}
v = \frac{F}{\gamma}\Big(
1 - \frac{p^2g^2}{2\gamma T}\int\!\!d{\bf r}\!\!\int_0^\infty \!\! dt\; 
\frac{\cos(pFt/\gamma)}{\big[ 1+ \Lambda^2(r^2 + Kt/\gamma)
\big]^{\eta}}\Big) \, ,
\label{vfcharPT3new}
\end{equation}
where we defined 
\begin{eqnarray}
\eta = \frac{Tp^2}{4\pi K}.
\end{eqnarray}
Taking the limit $F\to 0$ in the above expression, and performing the
time integration, we obtain
\begin{equation}
\lim_{F\to 0}(v/F) = \frac{1}{\gamma}\Big(
1 - \frac{\pi p^2g^2}{KT\Lambda^3}
\int_{0}^\infty\frac{dr}{( 1+ \Lambda^2 r^2)^{(2\eta -3)/2}}\Big),
\label{vF_limit}
\end{equation}
We now observe that the integral on the {\em rhs} of equation
(\ref{vF_limit}) behaves very differently depending on whether $T$ is
smaller or greater than
\begin{eqnarray}
T_{c0} = \frac{8\pi K}{p^2} \; .
\end{eqnarray}
For $T>T_{c0}$, {\em i.e.}, $\eta>2$, the integral in
Eq.~(\ref{vF_limit}) is convergent, and leads to a finite (and for
weak pinning $g$, to an arbitrarily small) correction to the linear
friction coefficient $\gamma(F=0)=1/\lim_{F\to 0}(v/F)$.  In strong
contrast, for $T<T_{c0}$ ($\eta<2$) above integral diverges signalling
the breakdown of the perturbation theory at small values of the
external force $F$.

\begin{center}
\begin{figure}[h]
\includegraphics[scale=0.8]{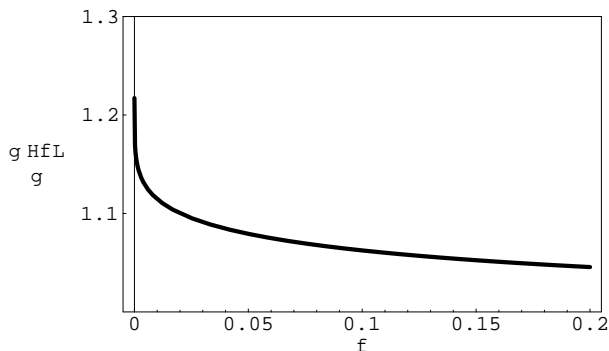}
\caption{
  Effective friction coefficient $\gamma(f)$ of the driven
  interface for $\eta=2.2$ and $(p^4g^2/8K^2\Lambda^4)=0.02$. As $f\to
  0$, $\gamma(f)$ remains finite, in agreement with equation
  (\ref{gamma_ggTc}).
}\label{gamma_rough}
\end{figure}
\end{center}

Having established the breakdown of perturbation theory for $T<T_{c0}$
in the limit of vanishingly small forces, we now turn our attention to
the full velocity-force characteristics at finite values of the
external drive. Starting from equation (\ref{vfcharPT3new}), and 
performing the integration over space variables, we obtain
\begin{eqnarray}
v=\frac{F}{\gamma}\Big(
1-\frac{p^4g^2}{8K^2\Lambda^4}\frac{\eta}{\eta-1}\int_0^\infty d\tau\;
\frac{\cos(2f\tau)}{(\tau+1)^{\eta-1}}
\Big) \; ,
\label{vfcharPT3}
\end{eqnarray}
where the dimensionless force $f$ is given by (henceforth, we shall
use both $F$ and $f$ to designate the driving force on our interface)
\begin{eqnarray}
f = \frac{pF}{2K\Lambda^2} \; . 
\label{def-f}
\end{eqnarray}
Performing the integral \cite{Prudnikov} on the {\em rhs} of equation
(\ref{vfcharPT3}), we finally arrive at the following result for the
effective friction coefficient $\gamma(f)$ of the driven interface
(here ${}_1F_2$ is a generalized hypergeometric function)
\end{multicols}
\begin{eqnarray}
\gamma(f)  =  \gamma\Big\{
1  -  \frac{p^4g^2}{8K^2\Lambda^4}\frac{\eta}{\eta-1}
\Big[ (2f)^{\eta-2}\Gamma(2-\eta)\sin\big(2f+\frac{\pi}{2}(\eta-1)\big) 
+ \frac{1}{(\eta-2)}\,{}_1F_2(1 ;\frac{4-\eta}{2},\frac{3-\eta }{2} ;-f^2)
\Big]
\Big\}^{-1} \; ,
\end{eqnarray}
which has the following limiting behavior as $f\to 0$,
\begin{eqnarray}
\gamma(f\to 0)  = \gamma\Big[
1  +  \frac{p^4g^2}{8K^2\Lambda^4}\frac{\eta}{(2-\eta)(\eta-1)}
\Big(1 - (2-\eta)\Gamma(2-\eta)\;(2f)^{\eta-2}\sin(\frac{\pi}{2}(\eta-1)\big)
\Big)\Big]^{-1} \; .
\label{vfcharPT4}
\end{eqnarray}
\begin{multicols}{2}
As found above, inside the rough phase, $T > T_{c0}$ ($\eta > 2$) and
for sufficiently weak pinning, the perturbation theory remains valid
at arbitrary $f$, simply displaying crossover from a freely moving
interface with ``bare'' mobility $\mu_\infty = 1/\gamma$ at high
drives to that with {\em finitely} suppressed low-drive mobility 
(as illustrated in figure \ref{gamma_rough})~:
\begin{eqnarray}
\gamma(f\to 0) \simeq \gamma\Big(1 + \frac{p^4g^2}{8\eta K^2\Lambda^4}\Big)
= \gamma\Big(1 + \frac{\pi p^2g^2}{2T K \Lambda^4}\Big).
\label{gamma_ggTc}
\end{eqnarray}
On the other hand, in agreement with Ref. \onlinecite{Blatter-et-al},
we find that in the ``smooth'', low temperature $T<T_c$ ($\eta<2$)
phase, the behavior is strikingly different with the correction to
$v_0(F)=F/\gamma$, Eq.~(\ref{vfcharPT4}), diverging and the
perturbative approach failing as $f$ is reduced below a characteristic
force
\begin{equation}
f_* (g,T) \approx \frac{1}{2}\,\left[
\frac{ \frac{p^4g^2}{8K^2\Lambda^4} 
\frac{
\eta \Gamma(2-\eta)}{(\eta-1)}
\sin(\frac{\pi}{2}(\eta-1)\big)
}
{1 + \frac{p^4g^2}{8K^2\Lambda^4}
\frac{\eta}{(2-\eta)(\eta-1)}
}\right]^{1/(2-\eta)} \; .
\label{fstarPT}
\end{equation}

\noindent
\begin{center}
\begin{figure}[h]
\includegraphics[scale=0.8]{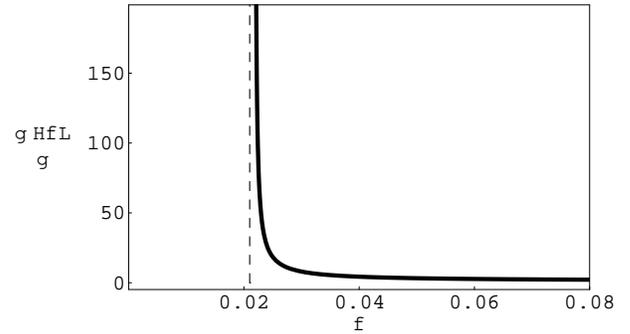}
\caption{
  Effective friction $\gamma(f)$ for $\eta=1.8$ and
  $(p^4g^2/8K^2\Lambda^4)=0.1$. $\gamma(f)$ diverges at
  $f_*\approx 0.022$ (dashed line), indicating the failure of perturbation
  theory at small drives $0<f<f_*$.
}\label{gamma<}
\end{figure}
\end{center}
\noindent
As $T\to T_{c0}^-$ ($\eta\to 2^-$),
\begin{eqnarray}
f_*(T_{c0}^-) \simeq \frac{1}{2}\,\exp\Big(-\frac{4K^2\Lambda^4}{p^4g^2}\Big),
\end{eqnarray}
showing that the regime of forces $0<f<f_*$ where perturbation theory
fails becomes exponentially small as $T_{c0}$ is approached from
below.  The unbounded growth of the perturbative friction coefficient
as the external drive $f$ approaches $f_*$ from above (see Fig.
\ref{gamma<}) suggests that the interface in the low temperature,
smooth phase is characterized by a vanishing linear mobility.
\cite{Nozieres,Blatter-et-al}

Although we will study this in more detail in following sections,
already at this stage we can see a physical interpretation of this
divergence. Perturbation theory in the pinning potential fails because
even for an arbitrarily weak pinning $g$, on sufficiently long scales
greater than $\xi$ (computed in Section \ref{DynamicRG}), the periodic
potential (for small $h$ acting like a ``mass'', $\frac{1}{2}g p^2
h^2$) necessarily dominates over the elastic energy density
$\frac{K}{2}\left(\nabla h\right)^2$.  Since (as is quite clear from
the equation of motion, Eq.~(\ref{eom_neq})) the applied force $F$
dominates the elastic force on scales longer than
\begin{equation}
\xi_F = \left(\frac{ 2\pi K}{p F}\right)^{1/2}\label{xiF},
\end{equation}
a sufficiently weak force, $F < F_*$, probes the interface on length
scales longer than $\xi$ and thereby leads to the breakdown of
perturbation theory about the harmonic interface.  Hence, although
quite instructive, the perturbation theory fails to make predictions
for $v(F)$ or any other dynamic quantity in the smooth phase at
sufficiently low drive $F<F_*$ and a nonperturbative approach is
necessary.

\section{Dynamic Renormalization Group}
\label{DynamicRG}

Armed with the above discussion, we are now well-equipped to use
dynamic RG analysis to make physical sense of these perturbative
divergences, with the main goal being the calculation of $v(F)$ in the
smooth $T<T_c$ phase for weak drive $F<F_*$.  It is convenient to
perform this analysis in the frame co-moving with the ``bare''
velocity $v_0=F/\gamma$ corresponding to the change of the dynamic
fields to $u({\bf r},t)\equiv h({\bf r},t)-Ft/\gamma$, which obeys
\begin{equation}
\gamma\,\partial_t u = K\nabla^2u + \frac{\lambda}{2}\,\big(\nabla u\big)^2
+pg\sin(pu+\frac{pF}{\gamma}t)  + \zeta({\bf r},t) \, .
\label{eq_motion2}
\end{equation}
Taking the nonlinear terms in the above equation as a small
perturbation, the equation of motion can be directly expanded in these
nonlinearities \cite{Halperin-et-al,Hohenberg,Ma} leading to
renormalization group recursion relations for model parameters. An
equivalent but more convenient formulation is the field-theoretic
approach of Martin, Siggia and Rose \cite{MSR} (MSR). In this
approach, the dynamic correlation and response functions
\begin{mathletters}
\begin{eqnarray}
C({\bf r},t)&=& \langle u({\bf r},t) u({\bf 0},0)\rangle \;,
\nonumber\\
&=&\int [du][d\tilde u]\,\,u({\bf r},t) u({\bf 0},0) \mbox{e}^{-S[u,\tilde u]},
\label{C}\\
R({\bf r},t)&=& \langle\tilde{u}({\bf r},t) u({\bf 0},0)\rangle \;,
\nonumber\\
&=&\int [du][d\tilde u]\,\,\tilde{u}({\bf r},t) u({\bf 0},0) 
\mbox{e}^{-S[u,\tilde u]},
\label{R}
\end{eqnarray}
\end{mathletters}
are computed directly by integrating over the phonon and response
fields $u$ and $\tilde u$, treated as independent stochastic fields
with a statistical weight $e^{-S[u,\tilde u]}$ imposed by the equation
of motion, after integrating over the thermal noise $\zeta({\bf r},t)$.  The
resulting effective ``action'' $S$ is given by $S = S_0+S_1$, where
\begin{eqnarray}
S_0[u,\tilde{u}] & = & \int  d{\bf r}dt\Big\{
\frac{1}{2}(2\gamma T)\tilde{u}^2({\bf r},t) +
\nonumber\\
& + & i\tilde{u}({\bf r},t)\big[
\,\gamma\partial_t u -K\nabla^2 u
\big]\Big\}
\end{eqnarray}   
is the action of a pinning-free (harmonic) interface, and where
$S_1=S_g+S_\lambda$, with
\begin{eqnarray}
S_g[u,\tilde{u}] = pg\;\int d{\bf r}dt \;i\tilde{u}({\bf r},t)\,\sin\big(pu({\bf r},t)+\frac{pFt}{\gamma}\big)
\label{s1gen}
\end{eqnarray}
the contribution of the pinning potential and 
\begin{eqnarray}
S_\lambda[u,\tilde{u}] = -\frac{\lambda}{2}\; \int d{\bf r}dt \; i\tilde{u}({\bf r},t)\,(\nabla u)^2
\end{eqnarray}
the contribution of the KPZ term to the nonlinearities in $S$. To
study the renormalization of $S[u,\tilde{u}]$ it is sufficient to work
with the dynamic ``partition function''
\begin{equation}
{\cal Z} = \int [du][d\tilde u]\;\mbox{e}^{-S[u,\tilde{u}]},
\end{equation}
required to remain fixed at unity under an RG coarse-graining procedure.
The advantage of the MSR formalism is its close resemblance to the
equilibrium statistical mechanics, that makes it a rather
straightforward task to apply RG transformations and to derive
recursion relations for the various parameters entering the equation
of motion (\ref{eq_motion2}).  Like the static momentum-shell RG, the
dynamic RG procedure consists of three main steps~:

{\em (i)} Thinning of the degrees of freedom, whereby modes $u({\bf
  q})$, with ${\bf q}$ in an infinitesimal shell $\Lambda/b<q<\Lambda$
($b=\mbox{e}^{d\ell}$) are perturbatively (in $S_1$) integrated out.

{\em (ii)} Rescaling of space variables according to ${\bf r}=b{\bf
  r}'$, so as to restore (for convenience) the ultraviolet cutoff to
its original value $\Lambda$, and rescaling time variable according to
$t=t'b^z$.

{\em (iii)} Rescaling of fields, in order (for convenience) to keep
the harmonic part of the action invariant under rescaling in {\it (ii)}.

We define ``slow'' $\{u^<,\tilde{u}^<\}$ and ``fast'' fields
$\{u^>,\tilde{u}^>\}$
\begin{eqnarray}
u({\bf q},t) & = & u^<({\bf q},t)+u^>({\bf q},t) \; ,
\\
u({\bf q},t) & = & u^<({\bf q},t)+u^>({\bf q},t) \; ,
\end{eqnarray}
with momentum support in Fourier space in the intervals
$0<q<\Lambda/b$ and $\Lambda/b<q<\Lambda$ respectively, and perform a
cumulant expansion of ${\cal Z}$ in terms of $S_1[u,\tilde{u}]$,
considered as a perturbation,
\begin{eqnarray}
{\cal Z} & = & 
\int [du][d\tilde u]\;\mbox{e}^{-S_0[u^<,\tilde{u}^<]}\Big\langle
\mbox{e}^{-S_1[u,\tilde{u}]}
\Big\rangle_{0>} \; ,
\nonumber\\
& \simeq & \int [du][d\tilde u]\;\mbox{e}^{-S_0[u^<,\tilde{u}^<] -\langle S_1\rangle_{0>} 
+ \frac{1}{2}\langle S_1^2\rangle_{0>}^c} \; ,
\end{eqnarray}
where $\langle\cdots\rangle_{0>}$ denotes an average taken with the
statistical weight $S_0[u^<,\tilde{u}^<]$, and where the superscript
$c$ in $\langle S_1^2\rangle_{0>}^c$ denotes a connected average.  To
first order in the pinning strength $g$, there is only one term in
$\langle S_g\rangle_0^>$, which renormalizes the dynamic action $S$,
namely
\begin{eqnarray}
\langle S_g\rangle_0^> &\equiv& pg b^{-Tp^2/4\pi K} \times
\nonumber\\
&\times&\int d{\bf r} dt \;i\tilde{u}^<({\bf r},t)
\sin\big(p u^<({\bf r},t)+\frac{pF}{\gamma}t\big),
\end{eqnarray}
which physically arises from the suppression (from $g$ to $g
b^{-Tp^2/4\pi K}$) of the effective pinning strength due to
short-scale thermal fluctuations averaging away the periodic
potential. In the above and throughout we will use $\equiv$ to indicates
that only the leading term has been kept.  Similarly, to first order
in the KPZ coupling $\lambda$, we have the following perturbative
correction to the dynamic action $S$,
\begin{eqnarray}
\langle S_\lambda\rangle_0^> &\equiv& 
-\int d{\bf r} dt \;i\tilde{u}^<({\bf r},t)
\left[\frac{\lambda T\Lambda^2}{4\pi K}\,d\ell\right],
\end{eqnarray}
which quite clearly renormalizes the effective external force.  

\medskip

Rescaling the space and time variables
\begin{mathletters}
\begin{eqnarray}
{\bf r} & = & b\,{\bf r}' \; , 
\label{space_rescaling}
\\
t & = & b^z\,t' \; ,
\label{time_rescaling}
\end{eqnarray}
\end{mathletters}
as well as the conjugate field $\tilde{u}({\bf r},t)$
\begin{eqnarray}
\tilde u\,^<({\bf r},t) & = & b^{\hat\chi}\tilde u\,({\bf r}',t') \; ,
\label{field_rescaling} 
\end{eqnarray}
while for convenience leaving $u({\bf r},t)$ unchanged in order to
preserve the periodicity $(2\pi/p)$ of the original
problem \cite{Remark3}
we obtain the following lowest-order
recursion relations~:
\begin{mathletters}
\begin{eqnarray}
(\gamma T)(b) & = & b^{2+z+2\hat\chi}\,(\gamma T) \; ,
\\
\gamma(b) & = & b^{2+\hat\chi}\,\gamma \; ,
\label{eq_gamma}
\\
K(b) & = & b^{z+\hat\chi}\, K \; ,
\label{eq_K}
\\
g(b) & = & b^{2+z+\hat\chi- Tp^2/4\pi K}\, g \; ,
\label{eq_g}
\\
\lambda(b) & = & b^{z+\hat\chi}\lambda \; ,
\label{eq_lambda}
\\
(F/\gamma)(b) & = & b^z(F/\gamma) \; .
\label{eq_F}
\end{eqnarray}
\end{mathletters}
The dynamic exponents $z$ and $\hat\chi$ can be fixed by requiring
that $K$ and $\gamma$ be unchanged, to linear order in $g$, under the
RG transformation.  This leads to the following values
\begin{eqnarray}
z = 2 \quad  , \quad \hat\chi =  -2 \nonumber
\end{eqnarray}
and to the following recursion relations for $g$ and $F$
\begin{mathletters}
\begin{eqnarray}
\frac{dg}{d\ell} & = & (2 - \frac{T p^2}{4\pi K}) \, g \; ,
\\
\frac{dF}{d\ell} & = & 2F + \frac{\lambda T\Lambda^2}{4\pi K} \; ,
\end{eqnarray}
\end{mathletters}
while the remaining quantities, $K$, $\gamma$, $\lambda$, and
temperature $T$, remain unchanged and suffer no renormalization to
first order in $g$ and $\lambda$.  Similar considerations, with
details given in Appendix B, lead to the following recursion relations
to second order in $g$ and~$\lambda$~:
\begin{mathletters}
\begin{eqnarray}
\frac{d}{d\ell}(\gamma T) & = & \big[\frac{T\lambda^2}{8\pi K^3}\! +
\!\frac{Tp^6g^2}{16\pi K^3\Lambda^4}\frac{1}{1+f^2}\big](\gamma T) \; ,
\label{rec2gamma}\\
\frac{d\gamma}{d\ell} & = & \frac{Tp^6g^2}{16\pi K^3\Lambda^4}\frac{1-f^2}{(1+f^2)^2}\,\gamma \; ,
\label{recgamma}\\
\frac{dg}{d\ell} & = & \big(2-\frac{Tp^2}{4\pi K}\big)\,g \; ,
\label{recg}\\
\frac{dK}{d\ell} & = & \frac{Tp^6g^2}{16\pi K^2\Lambda^4}\,\frac{2-3f^2-f^4}{(1+f^2)^3} \; ,
\label{recK}\\
\frac{d\lambda}{d\ell} & = & \frac{Tp^7g^2}{16\pi K^2\Lambda^4}\,\frac{f(f^2+5)}{(1+f^2)^3} \; ,
\label{reclambda} \\
\frac{dF}{d\ell} & = & 2F + \frac{\lambda T\Lambda^2}{4\pi K}-\frac{Tp^5g^2}{8\pi K^2\Lambda^2}\frac{f}{1+f^2} \; ,
\label{recF}
\end{eqnarray}
\end{mathletters}
where $f=(pF/2K\Lambda^2)$ is the dimensionless force of equation (\ref{def-f}).
Note that, because of the lack of a FDT for
the driven system, in strong contrast to the equilibrium case
($\lambda=F=0$) Eqs.(\ref{rec2gamma}) and (\ref{recgamma}) imply that
$T(\ell)$ flows nontrivially according to
\begin{eqnarray}  
\frac{dT}{d\ell} = \big[\frac{T\lambda^2}{8\pi K^3} +
\frac{Tp^6g^2}{8\pi K^3\Lambda^4}\;\frac{f^2}{(1+f^2)^2}\big]\;T \; .
\label{rec-T}
\end{eqnarray}
Hence, $T(\ell)$ is simply a measure of the strength of the
white-noise component of the random force on the driven interface and
is not associated with any equilibrium bath at a well-defined
thermodynamic temperature.

The recursion relations (\ref{rec2gamma})-(\ref{recF}) contain most
(but not all, as discussed in Section V) of the information we need to
investigate the properties of the system beyond the failing
perturbative expansion of Section \ref{secPT}. Before turning to their
full analysis and to the study of the velocity-force characteristics, it
is useful to see how the previously derived static and equilibrium
dynamic results \cite{Nozieres,Saito_Book} are recovered. We do this in
the following subsections.

\subsection{Analysis of the static limit}
\label{static_limit}

The static model, Eq.~(\ref{Hamiltonian}), is characterized by two
parameters $K$ and $g$ with the RG recursion relations reducing to the
familiar Kosterlitz-Thouless form (derived by these last authors in a
dual, Coulomb gas form\cite{KT})
\begin{mathletters}
\begin{eqnarray}
\frac{dg}{d\ell} & = & \big(2 - \frac{Tp^2}{4\pi K}\big) g \; ,
\label{recgsimple}\\
\frac{dK}{d\ell} & = & \frac{T p^6g^2 }{8 \pi K^2\Lambda^4} \; .
\label{recKsimple}
\end{eqnarray}
\end{mathletters}
At small $g$, $K(\ell)$ flows slowly, and the recursion relation for
$g$ implies the existence a phase transition (called ``roughening''
in the context of crystal surface \cite{Chui}) at $T_{c0}=8\pi
K/p^2$ (in the limit $g\to 0$) between two phases distinguished by the
long scale ($\ell\to\infty$) behavior of $g(\ell)$. For $T>T_{c0}$
thermal fluctuations are strong enough to effectively average away the
long-length scale effects of the periodic pinning potential, which is
therefore qualitatively unimportant for most (but not all) physical
properties of this so-called ``rough'' phase. At these high
temperatures the surface is logarithmically rough and the effects of a
weak periodic potential can be taken into account in a controlled
perturbative expansion. In strong contrast, for $T<T_{c0}$, the effective
strength of the periodic potential relative to that of the harmonic
elastic energy grows on long length scales, leading to a breakdown of
perturbation theory in $g$, no matter how weak its bare value might
be.  As a result, at long scales, the interface is pinned in this
``smooth'' phase, with bounded {\em rms} height fluctuations.

It is instructive to recall some of the physics which follows from the
above recursion relations.  It is convenient to first rewrite the flow
equations for dimensionless couplings $\tilde{g}$ and $\eta$
\begin{mathletters}
\begin{eqnarray}
\tilde{g} &=& \frac{\sqrt{2} p^2 g}{K\Lambda^2} \label{tilde-g} \; ,
\\
\eta &=& \frac{T p^2}{4\pi K} \; ,
\label{eta}
\end{eqnarray}
\end{mathletters}
that satisfy
\begin{eqnarray}
\frac{d\tilde g}{d\ell} & = & (2-\eta)\, \tilde{g} \; , \label{gen1}
\\
\frac{d\eta}{d\ell} & = & -\frac{1}{4}\eta^2 {\tilde{g}^2} \; . \label{gen2}
\end{eqnarray}

These show that in equilibrium, the quantity $\eta$ which is the measure of the
ratio of thermal ($T$) to elastic ($K$) energy, always flows to zero at
long scales, indicating that the low-temperature smooth phase is controlled
by a strong coupling zero-temperature fixed point. Near $T_c$, it is
convenient to use a reduced temperature 
measured relative to the (noninteracting) $T_{c0}=8\pi K/p^2$, 
\begin{mathletters}
\begin{eqnarray}
\tilde{\tau}&\equiv&\eta - 2\; ,\\
         &=&2 (T/T_{c0} - 1)  \; ,
\end{eqnarray}
\end{mathletters}
with the flow equations simplifying to
\begin{mathletters}
\begin{eqnarray}
\frac{d\tilde g}{d\ell} & = & -\tilde{\tau}\, \tilde{g} \; , \label{gen1Tc}
\\
\frac{d\tilde{\tau}}{d\ell} & = & - {\tilde{g}^2}. \label{gen2Tc}
\end{eqnarray}
\end{mathletters}
These can be easily integrated by multiplying Eq.~(\ref{gen1Tc}) and
(\ref{gen2Tc}) by $\tilde{g}$ and $\tilde{\tau}$, respectively, and
taking the difference of the two resulting equations.  The result is
that near $T_{c0}$ the flows are a family of hyperbolae
\begin{center}   
\begin{figure}[h]
\includegraphics[scale=0.8]{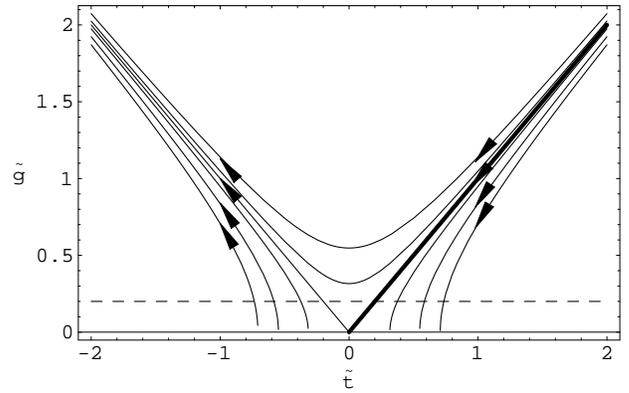}
\caption[Renormalization-group flow in the
$(\tilde{\tau},\tilde{g})$ plane for an interface driven in a periodic
potential]{ Renormalization-group flow in the
  $(\tilde{\tau},\tilde{g})$ plane. Temperature variation for an
  actual system occurs along the dashed line. On the high temperature
  side of the separatrix $\tilde{\tau}=\tilde{g}$ (indicated as the
  thick line), the periodic pinning $\tilde{g}$ renormalizes to zero
  and the interface is rough on long length scales. Below $T_c$ (to
  the left of the critical separatrix) the RG flow run off to strong
  coupling $\tilde{g}$ describing an interface that is smooth on long
  length scales.  }\label{fig_staticflow}
\end{figure}
\end{center}
\noindent

\begin{eqnarray}
\tilde{g}^2 - \tilde{\tau}^2 = c,
\label{hyperbola}
\end{eqnarray} 
labeled by a constant of integration
\begin{eqnarray}
c = \left(\frac{\sqrt{2} p^2 g}{K\Lambda^2}\right)^2 
- \left(\frac{T p^2}{4\pi K} - 2\right)^2,
\label{c}
\end{eqnarray}
determined by the bare value of model parameters $g$ and $K$.  The
resulting flows are illustrated in Fig. \ref{fig_staticflow}, showing
three distinct regions of behavior. In the high temperature region
below the thick line ($c < 0$), pinning is irrelevant, and it
therefore describes the rough phase, separated from the
low-temperature smooth phase (the region above the thick line) by a
critical line separatrix $\tilde{\tau}=\tilde{g}$.  The latter
therefore defines a true critical temperature given by
\begin{equation}
T_c=T_{c0}\left(1 + \frac{p^2 g}{\sqrt{2} K\Lambda^2}\right),
\end{equation}
distinct from its $g\to 0$ limit of $T_{c0}=8\pi K/p^2$.  Changing $T$
corresponds to the variation of the dimensionless bare parameters
along the dashed horizontal line indicated in Fig. \ref{fig_staticflow}. 

Above $T_c$, $\tilde{g}(\ell)$ flows to zero and 
\begin{mathletters}
\begin{eqnarray}
\tilde{\tau}_R &\equiv& \tilde{\tau}(\ell\to\infty),\\
  &=& \sqrt{|c|},
\label{t_R}
\end{eqnarray}
\end{mathletters}
corresponding to the long-scale renormalized elastic constant
\begin{mathletters}
\begin{eqnarray}
K_R &\equiv& K(\ell\to\infty),\\
    &=& K\frac{T}{T_{c0}}\left(1+\sqrt{|c|}/2\right)^{-1}.
\label{eqK_R}
\end{eqnarray}
\end{mathletters}
It is comforting to find (using Eq.~(\ref{c})) that $K_R$ reduces to its
bare value $K$ at high temperatures. Using the fact that near, but
above $T_c$

\begin{center}
\begin{figure}[h]
\includegraphics[scale=0.8]{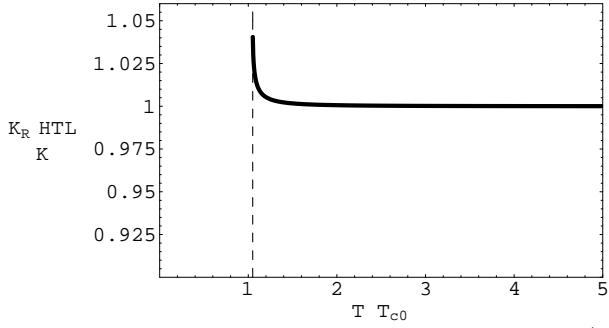}
\caption[Effective interface stiffness as a function of
  temperature for $\tilde{g}=0.1$]{
  Effective interface stiffness as a function of
  $T/T_{c0}$ for $\tilde{g}=0.1$.  In the smooth phase,
  $K_R$ scales with the system size, and is effectively infinite. At
  $T=T_c^+$, $K_R$ takes the value
  ${K}_R(T_c^+)=K(1+p^2g/\sqrt{2}K\Lambda^2)$ with a universal ratio
  $p^2/8\pi$ to the transition temperature $T_c$.  Far above $T_c$,
  $K_R$ goes to its bare value $K$. The dashed line indicates the location of
  $T_c$, which here is given by $T_c=1.05T_{c0}$.
}\label{fig_Kstatic}
\end{figure}
\end{center}
\noindent

\begin{mathletters}
\begin{eqnarray}
c &=& \tilde{g}^2 - (\tilde{\tau}_c + \tau)^2,\\
  &\approx & -2 \tilde{\tau}_c \tau,
\end{eqnarray}
\label{c_Tc}
\end{mathletters}
with the true reduced temperature relative to the true (finite $g$)
$T_c$ given by
\begin{eqnarray}
\tau &\equiv& \left(\frac{2T_c}{T_{c0}}\right)\frac{T-T_{c}}{T_{c}},
\end{eqnarray}
\label{t}
and $\tilde{\tau}_c = \tilde{g} = 2(T_c/T_{c0}-1)$, we find that in the limit
$T\to T_c^+$
\begin{eqnarray}
K_R(T) &=& K\frac{T_c}{T_{c0}}\left(1- \sqrt{\tilde{g}}
|T/T_{c}-1|^{1/2}\right).
\label{KR_Tc}
\end{eqnarray}
This leads at $T_c$ to a renormalized value of the elastic constant
$K_R(T_c^+)$ that is enhanced relative to the bare value $K$ and
with the universal ratio to $T_c$ given by 
\begin{equation}
\frac{K_R(T_c^+)}{T_c} = \frac{p^2}{8\pi} \; ,
\end{equation}
consistent with the analogous result first discovered in the context
of the $XY$-model \cite{NelsonKosterlitz}, related to our problem by
duality \cite{Jose,Villain-duality}.

Below $T_c$, the relative pinning strength runs off to strong coupling
and the interface is smooth on length scales longer than the
correlation length that we calculate below. Because the RG flows are
qualitatively very different near and away from the two separatrices
$\tilde{g}=\pm\tilde{\tau}$, the value of this important length scale,
that enters the velocity-force characteristics depends crucially on
the distance from $T_c$. In the critical region, defined by values of
the bare parameters such that the weak-coupling ($g$) flow is near and
roughly along either separatrix,
\begin{mathletters}
\begin{eqnarray}
\tilde{g}(\ell)&\approx&\pm\tilde{\tau}(\ell) \;,\\
&\approx&\frac{\tilde{g}}{1\pm\tilde{g}\ell},
\end{eqnarray}
\end{mathletters}
it is easy to show that the RG ``time'' $\ell_*$ to reach
strong-coupling is given by
\begin{mathletters}
\begin{eqnarray}
\ell_*&\approx&\frac{2}{\sqrt{c}}\;,\label{ellTc_a}\\
&\approx&\frac{2}{\sqrt{2\tilde{\tau}_c |\tau|}}\;.\label{ellTc_b}
\end{eqnarray}
\end{mathletters}
Consequently the correlation length in this critical region is of
familiar KT form \cite{KT}
\begin{mathletters}
\begin{eqnarray}
\xi_c &\approx& a\,e^{\ell_*},\label{xiTc_a}\\
&\approx& a\,e^{\alpha/|1-T/T_c|^{1/2}},\label{xiTc_b}
\end{eqnarray}
\end{mathletters}
diverging extremely fast as $T\to T_c^-$, with
$\alpha=\sqrt{2/\tilde{g}}=(T_c/T_{c0}-1)^{-1/2}$ a nonuniversal
constant.

Outside this critical region, defined by $\tau < -1$, deep in
the smooth phase, the flows are qualitatively different. At weak
coupling $g p^2\ll K\Lambda^2$ (the only regime where the perturbative
RG analysis is valid) because $\tilde{\tau}(\ell)$ grows weakly
(additively)
\begin{equation}
\tilde{g}(\ell)\approx\tilde{g} e^{(2-\eta)\ell},
\end{equation}
grows exponentially fast, reaching strong-coupling at the low-T
correlation length $\xi_g\approx a e^{\ell_g}$ given by
\begin{mathletters}
\begin{eqnarray}
\xi_g&\approx&\xi_0\left(\xi_0\Lambda\right)^{\eta/(2-\eta)},\\
&\approx&\Lambda^{-1}\big(\Lambda\xi_0\big)^{2/(2-\eta)},\\
&\approx&\Lambda^{-1}\left(\frac{K\Lambda}{p^2 g}\right)^{1/(2-\eta)}.
\label{xi_smooth}
\end{eqnarray}
\end{mathletters}

On scales longer than the roughness correlation length the interface
is smooth and is characterized by a strongly downward renormalized
value of the pinning strength $g_R$ determined by the value of
{\em  unrescaled} coupling $g(\ell=\log(\xi\Lambda))$ at the scale of the
correlation length. Near the transition 
\begin{mathletters}
\begin{eqnarray}
g_R&\approx& g(\Lambda\xi)^{-2}\ll g,\qquad T\to T_c^-,\\
   &\approx& g\,e^{-2\alpha/|1-T/T_c|^{1/2}}.
\end{eqnarray}
\label{gRcritical}
\end{mathletters}
Deep in the smooth phase, for weak pinning, we instead find
\begin{mathletters}
\begin{eqnarray}
g_R&\approx&g(\Lambda\xi)^{-\eta}\ll g,\qquad T\ll T_c \; ,\\
   &\sim& g^{2/(2-\eta)},
\end{eqnarray}
\label{gRsmooth}
\end{mathletters}
which for weak $g$ is also substantially reduced by thermal
fluctuations.

For strong pinning, fluctuations are unimportant and the correlation
length reduces to the substantially shorter strong-coupling value
$\xi_0 = (K/g p^2)^{1/2}$ determined by the bare model parameters.

\subsection{Analysis of the equilibrium dynamics}
\label{equilibrium_dynamics}

We now turn our attention to the equilibrium ($F=\lambda=0$) dynamics
of the sine-Gordon interface, characterized by an additional model
parameter, the friction coefficient $\gamma$, with the RG flow given
by
\begin{eqnarray}
\frac{d\gamma}{d\ell} & = & \frac{1}{8}\tilde{g}^2\eta\,\gamma \; .
\label{gammaflow_equil}
\end{eqnarray}
Combining this with the recursion relation Eq.~(\ref{recKsimple}), we
find that the renormalized surface stiffness $K_R$ and friction
coefficient $\gamma_R$ are related by
\begin{eqnarray}
\gamma_{R} = \gamma\,\Big(\frac{K_{R}}{K}\Big)^{1/2}.
\label{gamma_R}
\end{eqnarray}
This together with the results of the previous subsection, show that
the macroscopic {\em linear} mobility $\gamma^{-1}_R$ is finitely
renormalized in the rough phase, $T>T_c$ and displays a square-root
cusp approach to $\gamma_R^{-1}(T_c^+)=\gamma^{-1} (T_{c0}/T_c)^{1/2}$ as
$T\to T_c^+$
\begin{eqnarray}
\gamma_R^{-1}(T)&\approx& \gamma_R^{-1}(T_c^+)
\left(1 + \frac{1}{2}\sqrt{\tilde{g}}|T/T_{c}-1|^{1/2}\right),
\label{gammaR_Tc}
\end{eqnarray}
similar to the results of Petschek and Zippelius \cite{PetschekZippelius}
for the renormalized diffusion coefficient of the $XY$-model as $T\to T_{KT}^-$.

The effective friction coefficient $\gamma(\ell)$ at scale $e^\ell$
can be obtained by integrating the flow equation
(\ref{gammaflow_equil})
\begin{eqnarray}
\gamma(\ell) = \gamma\, 
\exp\left[\frac{1}{8}\int_0^{\ell}d\ell'\;\tilde{g}^2(\ell')\eta(\ell')\right].
\label{gamma_ell}
\end{eqnarray}
Since below $T_c$, at weak coupling, $\tilde{g}^2(\ell)\eta(\ell)$
grows with $\ell$, we find that the effective friction coefficient
runs off to infinity as $\ell\to\infty$ suggesting a vanishing of the
macroscopic linear mobility in the smooth phase. A more detailed
analysis of the equilibrium weak-coupling flow equations for large
$\ell$ gives
\begin{equation}
\gamma(\ell)\approx\gamma\left\{
\begin{array}{ll}
\exp\left[\frac{|\tau|\ell}{4\alpha^2}\right], \qquad   & T\to T_c^- \\
&\\
\exp\left[\frac{\eta\tilde{g}^2\,e^{2(2-\eta)\ell} }{16(2-\eta)}\right] . \qquad   
                             & T\ll T_c
\end{array}  
\right. 
\label{gamma_ell2}
\end{equation}
Such diverging friction coefficient can be physically interpreted as
activated creep dynamics over a pinning energy barrier that
asymptotically grows with length scale, logarithmically for $T\to
T_c^{-}$ and as a power-law for $T\ll T_c$. 

It is important to keep in mind that this growth of the friction
coefficient $\gamma(\ell)$ found in Eq.~(\ref{gamma_ell2}) extends only
up to the strong-coupling length scale $\xi=a e^{\ell_*}$ ($\xi_c$ for
$T\to T_c^{-}$, Eq. (\ref{xiTc_b}), and $\xi_g$ for $T\ll T_c$, Eq. (\ref{xi_smooth})) since it was 
derived based on a renormalization group approach that is perturbative in $\tilde{g}$.  
In Section \ref{Analysis}, we will look
in more detail at the physics on scales longer than $\xi$, but we can
already say at this point that (as we show in Section \ref{Analysis}) even in this
strong coupling regime the effective 
\begin{center}
\begin{figure}[h]
\includegraphics[scale=0.8]{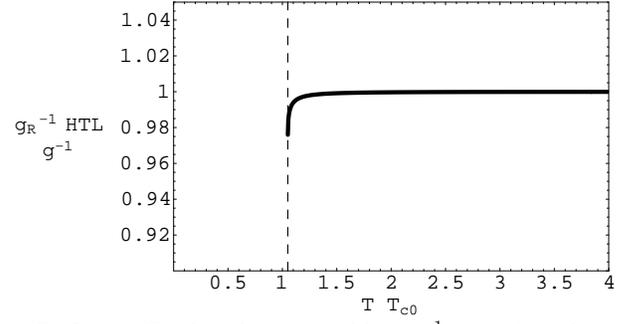}
\caption[Effective linear mobility $\gamma_{R}^{-1}$ as a
  function of temperature in equilibrium ($F=0$) for
  $\tilde{g}=0.1)$]{
  Effective linear mobility $\gamma_{R}^{-1}$ as a
  function of $T/T_{c0}$ in equilibrium ($F=0$) for
  $\tilde{g}=0.1$.  Below the roughening temperature at
  $T_c$, the mobility vanishes and the interface is pinned.
  $\gamma_{R}^{-1}(T)$ shows a square root cusp as $T\to T_c^+$, and
  goes to its bare value $\gamma^{-1}$ for $T\gg T_c$.
  The dashed line indicates the location of
  $T_c$, which here is given by $T_c=1.05T_{c0}$.
}\label{fig_mustatic}   
\end{figure}         
\end{center}
\noindent
friction coefficient diverges.
Consequently, we find that the interface linear (and in fact any
order-$n$) mobility exhibits a nonuniversal jump discontinuity to zero
across the roughening transition \cite{Nozieres,RostSpohn}, as
illustrated in Fig. \ref{fig_mustatic}.

\section{Non-equilibrium dynamics and the 
  velocity-force characteristics}
\label{Analysis}

\subsection{Weak-coupling regime}

We now turn to the full nonequilibrium problem, with the aim of
deriving the velocity-force characteristics of an interface driven through
a weak periodic potential, going beyond the failing (for $T<T_c$)
perturbative approach of section \ref{secPT}.  As long as the pinning
remains weak, the long scale physics of the driven interface is
contained in the renormalization group equations Eqs. 4.21-4.26, which
when rewritten in terms of the dimensionless variables $\tilde{g}$,
$\eta$, $f$ and the new KPZ coupling
\begin{eqnarray}
\tilde{\lambda} = \frac{\lambda}{pK}
\end{eqnarray}
are given by
\begin{eqnarray}
\frac{d\gamma}{d\ell} & = & \frac{1}{8}\,\eta\tilde{g}^2\,\frac{1-f^2}{(1+f^2)^2}\;\gamma \; ,
\label{rec-gamma}\\
\frac{d\tilde{g}}{d\ell} & = & ( 2 - \eta )\;\tilde{g} \; ,
\label{rec-tilde-g}\\
\frac{d\eta}{d\ell} & = & \frac{1}{2}\,\eta^2\tilde{\lambda}^2 + 
\frac{1}{8}\,\eta^2\tilde{g}^2\frac{-2 + 5f^2 + 3f^4}{(1+f^2)^3} \; , 
\label{rec-eta}\\ 
\frac{d\tilde{\lambda}}{d\ell} & = & \frac{1}{8}\,\eta\tilde{g}^2\frac{f(f^2+5)}{(1+f^2)^3} \; ,
\label{rec-lambda}\\
\frac{df}{d\ell} & = & 2f + \frac{1}{2}\,\eta\,\tilde{\lambda} - 
\frac{1}{8}\,\eta\tilde{g}^2\,\frac{f(3-f^2)}{(1+f^2)^3} \; .
\label{rec-f}
\end{eqnarray}
The most striking effect of nonequilibrium dynamics is the breakdown
of the FDT and as a result a nontrivial upward renormalization (flow)
of the effective ``temperature'' $T(\ell)$ driven by the external
force and the KPZ nonlinearity, reminiscent of nonequilibrium
``heating'' in randomly-pinned
systems \cite{KoshelevVinokur,Giamarchi3,BMR,Scheidl-Vinokur}.  Consequently,
even for $T<T_c$, for sufficiently strong drive the parameter
$2-\eta(\ell)$ determining the long-scale behavior of the periodic
potential is driven negative, leading to the irrelevance of the pinning
potential.  Hence, as discussed in the Introduction, a finite external
drive removes the qualitative distinction between the rough and smooth
phases and therefore rounds the roughening
transition. \cite{Nozieres,Hwa,RostSpohn,Remark1}

Here, we instead focus on the creep regime, where these particular
nonequilibrium effects are unimportant. In this weak driving creep
regime, we can ignore the KPZ nonlinearity and the most important role
of $F$, as can be clearly seen even at the level of perturbation
theory, Eq.~(\ref{vfcharPT}), and from the equation of motion, is to
introduce a new length scale $\xi_F\sim 1/\sqrt{F}$ defined in
Eq.~(\ref{xiF}).  Beyond this nonequilibrium length scale the effects
of the pinning potential and its ability to renormalize $\gamma(\ell)$
and $K(\ell)$ are suppressed, as it is averaged away on scales longer
than $\xi_F$ (see for example the RG flow equations above and analysis
below).  Hence, for weak external drive $F$, the effective values of
friction and interface stiffness parameters are given by
$\gamma(\ell_F)$ and $K(\ell_F)$ renormalized by Gaussian equilibrium
fluctuations up to length scale $\xi_F = e^{\ell_F}$.  This therefore
translates the strong $\ell$ dependence of $\gamma(\ell)$ into strong
$F$ dependence of the macroscopic mobility $\gamma^{-1}(F)$.
Substituting $\xi_F$, Eq.~(\ref{xiF}), inside our equilibrium flow,
Eqs.~(\ref{gamma_ell2}), and using
\begin{eqnarray}
v(F) = F/\gamma(\ell_F),
\label{v-F}
\end{eqnarray}
we immediately obtain the velocity-force characteristics,
Eq.~(\ref{vF>F*}), quoted in the Introduction.

This prediction for $v(F)$, Eq.~(\ref{vF>F*}), applies as long as the
relevant $F$ probes length scales $\xi_F$ on which the equilibrium
{\em weak-coupling} flow equations remain valid.  As discussed in the
previous section, these flows in fact break down due to strong
coupling effects (with $g$ itself cutting off thermal Gaussian
fluctuations) for length scales greater than $\xi$,
Eqs.(\ref{xiTc_b}),\ (\ref{xi_smooth}).  Hence, our predictions for
$v(F)$, Eq.~(\ref{vF>F*}), remain valid only as long as $\xi_F < \xi$
({\em i.e.}, it is the external force and not the periodic potential
itself that cuts off the Gaussian fluctuations), which translates into
the condition $F > F_*$, with the crossover force $F_*$ given by
equation ~(\ref{F*}) and in agreement with perturbation theory.

To see this weak-coupling phenomenology emerge directly from our full
nonequilibrium flow equations, Eqs.(\ref{rec-gamma})-(\ref{rec-f}), we integrate these
equations, with $\lambda=0$ and ignoring the nonequilibrium flow of
$T(\ell)$ (a valid approximation in the $F\to 0$ limit). We find for the renormalized friction coefficient 
the following intermediate result
\begin{eqnarray}
\gamma_{R}(f) = \gamma\,
\exp\left[\frac{1}{8}\int_0^{\infty}\!\!d\ell\,\eta(\ell)\,\tilde{g}^2(\ell)                    
\,\frac{1-f^2(\ell)}{\big(1+f^2(\ell)\big)^2}\right].
\label{gammaeff}
\end{eqnarray}
Since at low drive and weak coupling, well-below $T_c$, $\eta(\ell)$, $K(\ell)$,
and $T(\ell)$ grow slowly and $f(\ell)$ and $\tilde{g}(\ell)$ grow strongly
according to
\begin{mathletters}
\begin{eqnarray}
\tilde{g}(\ell)&=& \tilde{g}\, e^{(2-\eta)\ell}\;,\\
f(\ell)&\approx& f e^{2\ell},
\end{eqnarray}
\label{flow_simplified}
\end{mathletters}
it is quite clear from Eq.~(\ref{gammaeff}) that as long as the weak
coupling flows remain valid, in the smooth phase the flows are
automatically cut off when $f(\ell)$ gets to be $> 1$ leading to
$\ell_F$ discussed above.  

Substituting Eqs.~(\ref{flow_simplified}) into the expression of $\gamma_R(f)$, Eq.~(\ref{gammaeff}),
and integrating the resulting expression, we find
\begin{eqnarray}
\gamma_{R}(f) & = & \gamma\,\exp\Big(\frac{1}{8}\,\eta\tilde{g}^2\;A(-\frac{\tilde\tau}{2},f)\Big) \; ,
\label{gammaR}
\end{eqnarray}
with $-\frac{\tilde\tau}{2}=(2-\eta)/2=(1-T/T_{c0})$, and $A(x,f)$ is the function given by
\begin{eqnarray}
A(x,f) & = & \frac{1}{2f^4}\Big(\frac{1}{2(2-x)}
\,{}_2F^1\big(2,2-x,3-x,-f^{-2}\big)
\nonumber\\
&-& 2f^2(1-x)\,{}_2F^1\big(2,1-x,2-x,-\frac{1}{f^2}\big)\,\Big) ,
\label{function}
\end{eqnarray}

\begin{center}
\begin{figure}[h]

\includegraphics[scale=0.8]{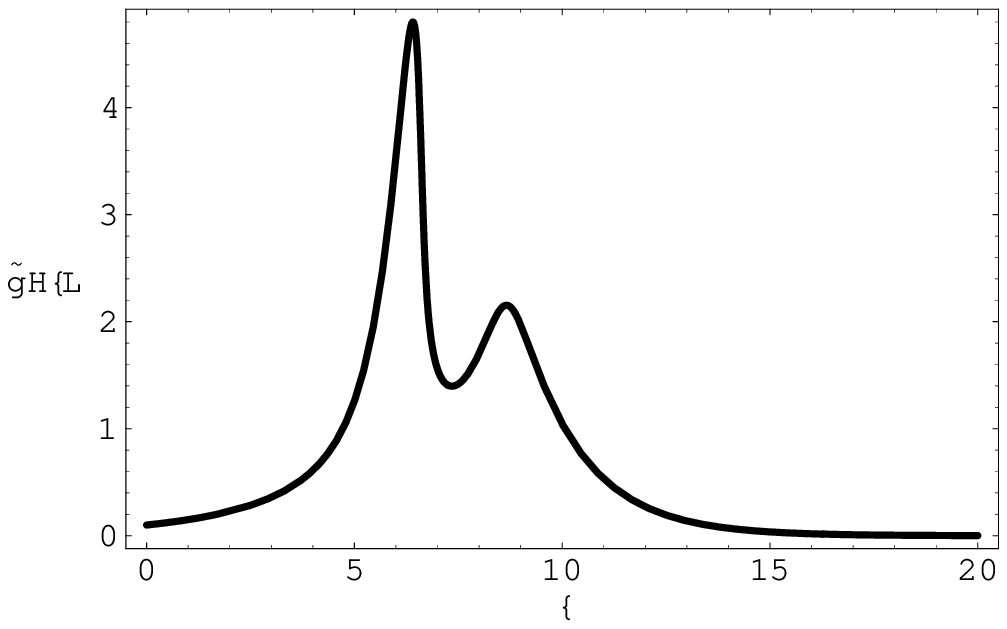}

\includegraphics[scale=0.8]{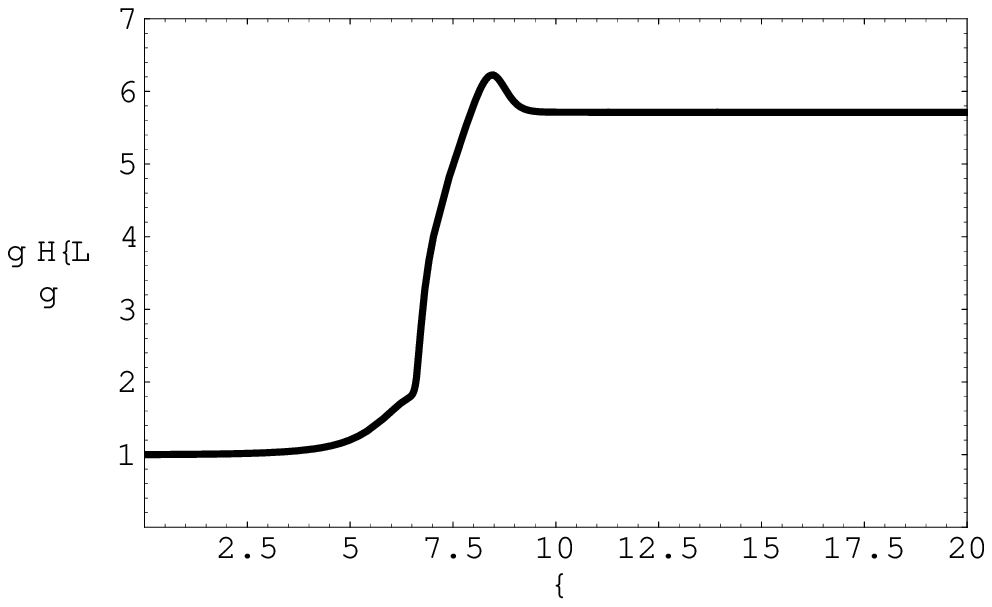}
\caption{
  Behaviour of the pinning strength $g$ (top) and of the
  friction coefficient $\gamma(\ell)$ (bottom) with length scale
  $\ell$ for $\tilde{g}=0.1$, $T\simeq 0.8T_c$ and
  $f\simeq 1.7374\times 10^{-5}$.  Here $f_*\simeq 1.7373\times 10^{-5}$.    
}\label{fig_gtozero}
\end{figure}
\end{center}

\begin{center}
\begin{figure}[h]
\includegraphics[scale=0.8]{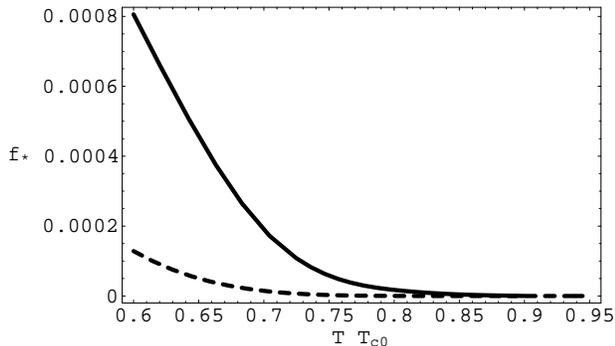}
\caption[Characteristic force $f_*(T)$ as obtained from the numerical
  solution of the dynamic RG recursion relations (solid line) and from the
  perturbative estimate of equation (\ref{fstarPT}), (dashed line),
  for $\tilde{g}=0.1$]{
  Characteristic force $f_*(T)$ as obtained from the numerical
  solution of the dynamic RG recursion relations (solid line) and from the
  perturbative estimate of equation (\ref{fstarPT}), (dashed line),
  for $\tilde{g}=0.1$. The curve $f_*(T)$ delimits two
  very different physical regimes. Above this curve, the interface moves
  with uniform velocity. On the other hand, for $f<f_*(T)$, the interface
  moves through the nucleation of soliton excitations.
}\label{fig_fstar}
\end{figure}
\end{center}
\noindent
where ${}_2F^1$ denotes a hypergeometric function \cite{Abramowitz}.
When $T<T_c$ ({\em i.e.} $(-\tilde\tau)>0$), taking the limit of the
function $A(-\frac{\tilde\tau}{2},f)$ of equation (\ref{gammaR}) when $f\to 0$
leads to the following expression for the long-scale inverse 
of nonlinear mobility $\gamma_{R}$
\begin{eqnarray}
\gamma_{R}(f) = \gamma \,\exp\big( (F_*/F)^{2(1-T/T_{c})}\big),
\label{result1_belowTr}
\end{eqnarray}   
with
\begin{eqnarray}
F_*(\tilde{g},T) \simeq 
\frac{2K\Lambda^2}{p}\Big(\frac{\eta\tilde{g}^2}{16(2-\eta)}\Big)^{\frac{1}{2(1-T/T_c)}}\; ,
\label{estimate-Fstar}
\end{eqnarray}
in full agreement with earlier more qualitative discussion of the
velocity-force characteristics in the intermediate regime of forces
$F>F_*$, and $F_*$ consistent with the perturbative result
(\ref{fstarPT}) for $\tilde{g}\ll 1$.

As $F$ is lowered below $F_*$, eventually the saturation of
$\gamma(\ell)$ breaks down and the flow behavior changes dramatically
as strong-coupling length scales (at which our weak-coupling RG
solution is invalid) are probed. Studying the point at which this
happens as a function of model parameters, allows us to extract the
crossover value of $F_*$, which we plot in Fig. \ref{fig_fstar}.  We
find that there is a qualitative agreement between the analytical
prediction for $f_*$, Eq. (\ref{fstarPT}), and our numerical analysis.

\subsection{Strong-coupling regime}

The weak-coupling behavior found in the previous subsection only
extends up to the scale $\xi$, Eqs.(\ref{xiTc_b}),\ (\ref{xi_smooth}).
Beyond this strong-coupling length, in the equilibrium model, the
growth of $\tilde{g}(\ell)$ and $\gamma(\ell)$ is cutoff by the
pinning potential, and an approach nonperturbative in $\tilde{g}$,
where pinning is treated on equal footing with the elastic energy, is
required.  In this strong-coupling regime Gaussian interface
fluctuations, considered so far, are strongly suppressed by

\begin{center}   
\begin{figure}[h]
\includegraphics[scale=0.5]{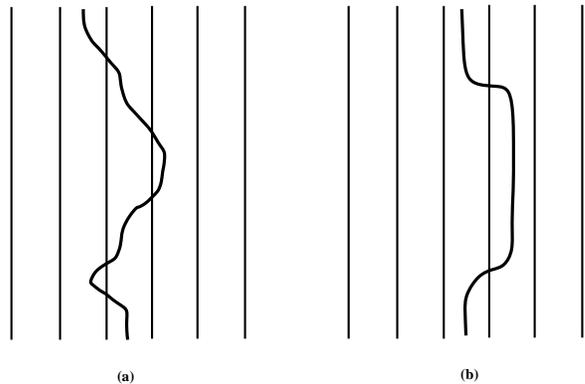}
\caption[Schematic representation of the motion of a driven
  interface past the periodic pinning potential]{
  Schematic representation of the motion of a driven
  interface past the periodic pinning potential. (a) When $T>T_c$, or  
  $T<T_c$ and $f>f_*$, the large fluctuations of the interface wash
  out the pinning potential on large length scales and the interface
  moves with a uniform velocity. (b) On the other hand, for
  $T<T_c$ and $f<f_*$ the fluctuations of the interface are small;
  as a result, most of the interface is pinned at a given minimum 
  of the pinning potential, and motion from one minimum to the next
  takes place through soliton excitations.
}\label{fig_physics_motion}
\end{figure}
\end{center}
\noindent
the pinning barrier that scales like $L^2$ relative to the elastic
energy.

Instead, at low temperature the fluctuations are dominated by
nontrivial saddle-point solutions (solitons) of $H$,
Eq.~(\ref{Hamiltonian}), with model parameters, $K_R$, $g_R$,
$\gamma_R$ renormalized by Gaussian fluctuations on weak-coupling
scales $L<\xi$.  The dominant soliton excitation, illustrated in
projection in Fig. \ref{fig_physics_motion}, corresponds to a circular
patch of radius $R>\xi$ of a nearly flat interface moving over to a
neighboring minimum of the periodic potential, with an energy cost
that clearly grows linearly with $R$
\begin{equation}
E_{\rm soliton}(R)\approx p\,g_R \xi R\label{Esoliton} \; ,
\end{equation}
where $g_R$ (Eqs. (\ref{gRcritical}), (\ref{gRsmooth})) and $\xi$
(Eqs.(\ref{xiTc_b}), (\ref{xi_smooth})) strongly depend on the proximity to
$T_c$. At zero-drive, the barrier to such solitonic motion simply
diverges and linear mobility vanishes identically. A velocity-force
characteristics in the weak drive $F < F_*$ ({\em i.e.}, $\xi < \xi_F$)
regime can be analyzed via scaling nucleation
theory \cite{Nozieres}.  In this creep regime the interface is
in near metastable equilibrium with $F$ introducing a contribution
\begin{equation}
E_{\rm F}(R)\approx -F\, \pi R^2 d \label{E_F}
\end{equation}
to the effective free energy. Balancing $E_F(R)$ against the soliton
energy $E_{\rm soliton}(R)$ we find that solitons of size larger than
a critical radius
\begin{equation}
R_{\rm c}\approx \left(\frac{p\,g_R \xi}{2\pi d}\right)\frac{1}{F}\label{Rc}
\end{equation}
are unstable. In the $F\to 0$ limit, thermal activation rate of solitons
of size $R_c\sim 1/F$ is quite clearly the limiting step for interface
creep motion. We therefore find that the weak-coupling velocity-force
characteristics, Eq.~(\ref{vF>F*}) crosses over, for $F<F_*$, to that
given by Eq.~(\ref{vF<F*}) in the Introduction, with
\begin{eqnarray}
F_0 \approx \frac{(pg_R\xi)^2}{4\pi d} \; .
\label{F_0}
\end{eqnarray}
For vanishing temperature and strong bare pinning potential our asymptotic (for
$F<F_*$) result for $v(F)$ reduces to that found in
Refs. \onlinecite{Nozieres,Blatter-et-al}. However, at large $T < T_c$
and weak bare pinning $g$, we predict a strong thermal renormalization
of the characteristic pinning energy
\begin{equation}
p^2 g^2 a^2\to p^2 g_R^2 \xi^2
\end{equation} 
by thermal fluctuations on scales smaller than $\xi$.

\section{Discussion and Conclusions}
\label{sine_G_Conclusion}

In this paper we have studied the creep dynamics of a two-dimensional
interface driven through a periodic potential. Using dynamic
renormalization group methods and matching to strong coupling, we have
calculated the velocity-force characteristics across the interface
roughening transition. Consistent with previous studies, we find a
qualitative change across the transition in the weak-drive
velocity-force characteristics, with Ohmic transport for $T>T_c$ and 
a jump discontinuity in mobility across the transition. For $T<T_c$, in
the asymptotic creep regime $F\ll F_*(g,T)$ and for strong bare
coupling, where transport is via soliton activation at all scales, we
recover previously found results for the velocity-force characteristics $v(F)$.  
However, for weak bare coupling and
strong thermal fluctuations, we predict an intermediate drive $F >
F_*(g,T)$ nonlinear regime with a continuously varying (with $T$)
exponent, which asymptotically crosses over to the strong coupling
result with strongly thermally renormalized characteristic pinning
barrier.  Unfortunately, because the characteristic force $F_*(g,T)$
that delineates between the intermediate drive regime and the
strong-coupling regime coincides with the force marking the breakdown
of the perturbative high-velocity expansion, we expect it to be
difficult to observe this intermediate drive regime.

The new physical picture which emerges from the present study
complements previously made predictions \cite{Blatter-et-al} which were
based on a more elementary perturbative approach, as well as known
results for the mobility \cite{Nozieres,RostSpohn} at zero external
drive. On the experimental side, the above picture may shed some light
on experiments such as those of Wolf et {\em al.} \cite{Wolf-et-al},
who found that the growth velocity $v$ of a surface of crystalline
Helium 4 is strongly reduced at $T_c$ from an Ohmic behavior $v\sim
F$ for $T>T_c$ to an extremely slow growth rate for $T<T_c$, a result
which is usually explained in terms of an onset of creep motion via
soliton like excitations \cite{Nozieres}.

An interesting and experimentally relevant generalization of our
results is a study of creep dynamics of a two-dimensional solid,
driven through a one- or two-dimensional periodic potential, with
applications to driven 2D colloidal crystals and vortices in
superconducting films. Despite of considerably different geometry,
in equilibrium these systems display a
pinned-to-floating solid transition closely related to the roughening
transition of 2D interfaces. However, new interesting ingredients
arise. Some of the most important ones are the nonequilibrium
conventive-like terms \cite{ChaFertig,BalentsFisher,BMR}, vector phonon
displacement and concomitant possible importance of dislocations.
Combined with the considerably interesting behavior of the scalar
sine-Gordon model studied here, we expect these to lead to even richer
phenomenology. We expect that studies of these will shed considerable
light on numerous experiments 
and simulations \cite{simulations}.

\acknowledgments 

L.R. thanks Harvard Physics Department for hospitality and Daniel
Fisher and John Toner for discussions. We acknowledge financial
support by the National Science Foundation through grant DMR-9625111,
and by the A.  P. Sloan, and David and Lucile Packard Foundations.

\section{Appendix~: Static momentum shell renormalization group}

In this Appendix, we present technical details on the derivation of
the renormalization group recursion relations for the driven
sine-Gordon model in $2+1$ dimensions. For completeness, we shall
begin in section \ref{App_staticRG} by showing how the standard
momentum shell\cite{Wilson-Kogut} RG with hard cutoff
\cite{Wegner-Houghton,Kogut} can be applied to the static version of
this problem before deriving the full dynamic equations at nonzero
external drive in section \ref{App_sineG_B}.

\subsection{Static RG}
\label{App_staticRG}

We decompose the field $h({\bf r})$ in the Hamiltonian 
(\ref{Hamiltonian}) into high and low wavevector components
\begin{eqnarray}
h({\bf r}) = h^<({\bf r}) + h^>({\bf r})
\end{eqnarray}
such that
\begin{eqnarray}
h^<({\bf r}) & = & \int_{\bf q}^< h({\bf q})\,\mbox{e}^{i{\bf q}\cdot{\bf r}} \; ,
\\
h^>({\bf r}) & = & \int_{\bf q}^> h({\bf q})\,\mbox{e}^{i{\bf q}\cdot{\bf r}} \; ,
\end{eqnarray}
where $\int_{\bf q}^<\equiv \int_0^{\Lambda/b}\frac{d\bf q}{(2\pi)^2}$ and 
$\int_{\bf q}^>\equiv \int_{\Lambda/b}^{\Lambda}\frac{d\bf q}{(2\pi)^2}$  
denote integration in momentum space over the 
ranges $0<|{\bf q}|<\Lambda/b$ and $\Lambda/b<|{\bf q}|<\Lambda$ respectively. 
In terms of these high and low momentum fields, the equilibrium Hamiltonian 
$H_0[h]=\frac{1}{2}\int d{\bf r}\, K\big(\nabla h\big)^2$
can be written as the sum
$$H_0[h]=H_0[h^<]+H_0[h^>]\; .$$ 
We now want to integrate over the fast component $h^>({\bf r})$. To this end, we rewrite
the partition function $Z =\int [dh]\exp(-\beta H)$ in the form (here $\beta=1/T$ is the inverse 
temperature)
\begin{eqnarray}
Z & = & \int [dh^<][dh^>]\mbox{e}^{-\beta H_0[h^<]-\beta H_0[h^>]- \beta H_1[h^<+h^>]} \; ,
\nonumber \\
& = & \int [dh^<] \mbox{e}^{-\beta H_0[h^<]} \int [dh^>]\mbox{e}^{-\beta H_0[h^>]-\beta H_1[h^<+h^>]} \; , 
\nonumber \\
& = & \int [dh^<] \mbox{e}^{-\beta H_0[h^<]+\beta \ln Z_0^>}
\big\langle\mbox{e}^{-\beta H_1[h^<+h^>]}\big\rangle_{0>} \; ,
\label{eqZstat}
\end{eqnarray}
where $Z_0^>=\int [dh^>]\,\exp(-\beta H_0[h^>])$, and where the subscript $(0>)$ means that the average
with respect to $h^>$ is performed with statistical weight $\exp(-\beta H_0[h^>])/Z_0^>$. The term between 
angular brackets in Eq.~(\ref{eqZstat}) is then approximated by a cumulant expansion
\begin{equation}
\big\langle\mbox{e}^{-\beta H_1[h^<+h^>]}\big\rangle_{0>} = 1 - \frac{\langle H_1\rangle_{0>}}{T} +
\frac{1}{2T^2} \langle H_1^2\rangle_{0>}^c + \cdots
\label{cumul1}
\end{equation}
where $\langle H_1^2\rangle_{0>}^c$ denotes the second cumulant
$\big\langle (H_1^2-\langle H_1\rangle^2)\big\rangle_{0>}$.
When re-exponentiated, equation (\ref{cumul1}) leads to the result
\begin{equation}
\big\langle\mbox{e}^{-\beta H_1[h^<+h^>]}\big\rangle_{0>} = \mbox{e}^{-\beta H_{eff}} ,
\end{equation}
with the effective Hamiltonian
\begin{eqnarray}
H_{eff} = \langle H_1\rangle_{0>} - \frac{1}{2T} \langle H_1^2\rangle_{0>}^c + \cdots
\end{eqnarray}
The averages in Eq.~(\ref{cumul1}) can be easily evaluated, with the results\cite{Ohta,Nozieres}
\end{multicols}
\begin{eqnarray}
\langle H_1\rangle_{0>} & = & -g\mbox{e}^{-\frac{1}{2}p^2G^>(0)}\int d{\bf r}\,\cos[ph^<({\bf r})] \; ,
\label{statcumul1}\\
\langle H_1^2\rangle_{0>}^c & = & \frac{1}{2}g^2\mbox{e}^{-p^2G^>(0)}
\int d{\bf r}\,d{\bf r'}\,\big[\mbox{e}^{p^2G^>({\bf r}-{\bf r'})}-1\big]
\Big\{
\cos\big(p(h^<({\bf r})+h^<({\bf r'}))\big) + \cos\big(p(h^<({\bf r})-h^<({\bf r'}))\big)\,\Big\} \; ,
\label{statcumul2}
\end{eqnarray}
\begin{multicols}{2}
\noindent where $G^>({\bf r}-{\bf r}')=\langle h^>({\bf r})h^>({\bf r}')\rangle_{0>}$ is the elastic propagator 
for fast fields (here $J_0$ is the zeroth order Bessel function)
\begin{equation}
G^>({\bf r}-{\bf r'}) = T\int_{\bf q}^>\frac{\mbox{e}^{i{\bf q}\cdot({\bf r}-{\bf r'})} }{Kq^2}
=\frac{T\,d\ell}{2\pi K}\;J_0(\Lambda|{\bf r}-{\bf r'}|) \; .
\label{statcorr}
\end{equation}
Given that $G^>(0)=Td\ell/2{\pi}K$, we see that the first order cumulant (\ref{statcumul1}), after the rescalings
(\ref{rescaling1})-(\ref{rescaling2}), 
leads straightforwardly to the recursion relation (\ref{recgsimple}) for the pinning strength  $g$.
On the other hand, since the ``kernel''
\begin{equation}
{\cal K}({\bf r})=\big[\mbox{e}^{p^2G^>({\bf r})}-1\big]\label{kernel}
\end{equation}
takes appreciable values only for small values of its argument, we see that
the first term inside the integral in equation (\ref{statcumul2}) will contribute higher harmonic terms 
($\sim\cos\big(2ph({\bf r})\big)$) to the effective Hamiltonian, and hence we shall discard this term as
irrelevant. In the second term, we shall make the approximation 
\begin{eqnarray}
\cos\big[\!&p&\!\big(h^<({\bf r})  \!-\!  h^<({\bf r'})\big)\big] \simeq 1 -\frac{1}{2}p^2\big(h^<({\bf r})-h^<({\bf r'})\big)^2 \; ,
\nonumber\\
& \simeq & 1 -\frac{1}{2}p^2({\bf r}-{\bf r'})_\alpha
({\bf r} \! - \!{\bf r'})_\beta\partial_\alpha h^<({\bf r})\partial_\beta h^<({\bf r})  ,
\label{approxhhp}
\end{eqnarray}
where, in going from the first to the second line, we made use of the Taylor expansion
$$h^<({\bf r})-h^<({\bf r'})\simeq ({\bf r} - {\bf r'})_\alpha\partial_\alpha h^<({\bf r})\, .$$
Inserting (\ref{approxhhp}) back into Eq.~(\ref{statcumul2}), we obtain the
following approximation to the second cumulant 
(we here use the symbol $\equiv$ to indicate that we retain only the term
correcting the stiffness $K$)
\begin{eqnarray}
\langle H_1^2\rangle_{0>}^c & \equiv & -\frac{1}{8}\,p^2g^2\mbox{e}^{-p^2G^>(0)}
\int d{\bf r}\;\big(\nabla h^<({\bf r})\big)^2\times\nonumber\\
&\times& \int d{\bf r'}\, ({\bf r}-{\bf r'})^2{\cal K}({\bf r}-{\bf r'}) \; .
\label{2ndcumul}
\end{eqnarray}
Since $G^>\propto d\ell=\ln b$, we can expand the exponential in a Taylor series in $G^>$,
\begin{equation}
\mbox{e}^{p^2G^>({\bf r}-{\bf r'})}-1\simeq p^2G^>({\bf r}-{\bf r'}) + 
\frac{1}{2!}p^4\big(G({\bf r}-{\bf r}')\big)^2 \; .
\label{expansion1}
\end{equation}
Now, the renormalization of $K$ involves the integral
\begin{eqnarray}
\int d{\bf r}\; {\bf r}^2 \,{\cal K}({\bf r}) 
= \int d{\bf r}\;{\bf r}^2\big[\mbox{e}^{p^2G^>({\bf r})}-1\big] \; .
\label{eqintI}
\end{eqnarray}
Inserting the expansion (\ref{expansion1}) into this last expression, 
the first term gives a contribution
\begin{eqnarray}
\int d{\bf r}\;r^2G^>({\bf r}) = - \nabla_{\bf q}^2G^>({\bf q})\big|_{q=0}
\label{vanishes}
\end{eqnarray}
which vanishes identically, since $G^>({\bf q})$ has support only on the shell $\Lambda/b<q<\Lambda$.
The second term gives
\begin{eqnarray}
\int d{\bf r}\; {\bf r}^2 \,{\cal K}({\bf r}) & = & \frac{1}{4}p^4\int d{\bf r} \; r^2\big(G^>({\bf r})\big)^2  
\nonumber\\
& = & \frac{T^2p^4\ln b}{\pi K^2\Lambda^4} \; .
\label{int-rsqrd}
\end{eqnarray}
Thus, we obtain for the second cumulant (\ref{2ndcumul}) the following expression
\begin{eqnarray}
-\frac{\langle H_1^2\rangle_{0>}^c}{2T} & \equiv & \frac{Tp^6g^2 \;d\ell}{8\pi K^2\Lambda^4}
\int d{\bf r}\;\frac{1}{2}\big(\nabla h^<({\bf r})\big)^2 \;.
\nonumber
\end{eqnarray}
We now perform the following rescalings
\begin{eqnarray}
{\bf r} & = & \mbox{e}^{\ell}\,{\bf r}' \; ,
\label{rescaling1}
\\
h^<({\bf r}) & = & \mbox{e}^{\chi\ell}\, h({\bf r}') \; ,
\label{rescaling2}
\end{eqnarray}
so as to restore the ultraviolet cut-off back to $\Lambda$.  Because
the pinning potential is a periodic function, it is convenient
(although not necessary) to set the arbitrary field dimension $\chi$
to zero, thereby preserving the period $2\pi/p$ of the original
problem under RG transformations. Under such a transformation, the
resulting effective Hamiltonian can be cast into its original form
with effective $\ell$-dependent parameters $K(\ell)$ and $g(\ell)$
such that
\begin{mathletters}
\begin{eqnarray}
g(\ell) & = & g \mbox{e}^{(2-Tp^2/4\pi K)\ell} \; ,
\\
K(\ell) & = & K +\frac{Tp^6g^2}{8\pi K^2\Lambda^4}\;d\ell \; ,
\end{eqnarray}
\end{mathletters}
or, in differential form
\begin{mathletters}
\begin{eqnarray}
\frac{dg}{d\ell} & = &  \Big(2- \frac{Tp^2}{4\pi K}\Big)\, g \; ,
\\
\frac{dK}{d\ell} & = &  \frac{Tp^6g^2}{8\pi K^2\Lambda^4} \; .
\end{eqnarray}
\end{mathletters}

\subsection{Dynamic RG}
\label{App_sineG_B}

We now turn our attention to the derivation of the dynamic RG flow equations
(\ref{rec2gamma})-(\ref{recF}) for the driven sine-Gordon model. As we did in the static case, 
we define the following low and high momentum components $h^<({\bf r},t)$ and $h^>({\bf r},t)$ :
\begin{eqnarray}
u^<({\bf r},t)= \int_{{\bf q},\omega}^< h({\bf q},\omega)\,\mbox{e}^{i({\bf q}\cdot{\bf r}-\omega t)} ,
\\
u^>({\bf r},t)= \int_{{\bf q},\omega}^> h({\bf q},\omega)\,\mbox{e}^{i({\bf q}\cdot{\bf r}-\omega t)} ,
\end{eqnarray}
where, here and in what follows, $\int_{{\bf q},\omega}$ on integrals
stands for $\int \frac{d^d{\bf q}}{(2\pi)^d}\frac{d\omega}{2\pi}$, and
the superscripts $>$ and $<$ indicate integration over the high
($\Lambda/b < q < \Lambda$) and low ($0<q<\Lambda$) momentum regions,
respectively.  Using the fact that $u({\bf q},\omega)=u^<({\bf
  q},\omega)+u^>({\bf q},\omega)$, it is not difficult to verify that
the free part $S_0$ of the action decomposes into two diagonal pieces
$S_0^<$ and $S_0^>$ depending only on $u^<({\bf q},\omega)$ and
$u^>({\bf q},\omega)$ respectively
\begin{eqnarray}
S_0[u,\tilde u] = S_0[u^<,\tilde u^<]  + S_0[u^>,\tilde u^>] \; .
\end{eqnarray} 
As we did in the static RG, in order to be able to integrate out the
fast component of the field $u({\bf r},t)$, we rewrite the generating
functional ${\cal Z}$ in the form
\begin{eqnarray}
{\cal Z} 
& = & \!\int \! [du^<][d\tilde u^<]\mbox{e}^{-S_0[u^<,\tilde u^<]+\ln {\cal Z}_0^>}
\langle \mbox{e}^{-S_1[ u^<+ u^>, u^<+ u^>]} \rangle_0^>
\nonumber
\end{eqnarray} 
where ${\cal Z}_0^>=\int [du^>][d\tilde u^>]\,\,\exp(-S_0[ u^>,\tilde u^>])$, and where 
$\langle \cdots\rangle_0^>$ denotes statistical averaging with statistical weight 
$\mbox{e}^{-S_0[u^>,\tilde u^>]}$. The perturbative correction to the dynamic action can therefore
be expressed in terms of a cumulant expansion
\begin{eqnarray}
\langle \mbox{e}^{-S_1}\rangle_{0>} = 1 - \langle S_1\rangle_{0>} +
\frac{1}{2}\,\big\langle S_1^2\big\rangle_{0>} +\cdots
\label{cumul} 
\end{eqnarray}
Reexponentiation of this expansion allows us to define the effective action
\begin{equation}
S_{eff}[u,\tilde{u}] = S_0 + \langle S_1\rangle_{0>} -
\frac{1}{2}\,\Big\langle\big[S_1^2-\big(\langle
S_1\rangle_0^>\big)^2\big]\Big\rangle_{0>} +\cdots 
\label{cumulant}
\end{equation} 
from which we can derive dynamic RG flows for the parameters of the
original equation of motion.  This procedure, to first order in the
pinning strength $g$, has already been shown in the text.  Here we are
therefore only going to consider the second order correction to the
original action $S$.  In fact, it turns out\cite{RostSpohn} that the
only perturbative corrections to $S$ to second order in perturbation
theory come from the cumulants $-\frac{1}{2}\langle S_g^2\rangle_{0>}$
and $-\frac{1}{2}\langle S_\lambda^2\rangle_{0>}$, {\em i.e.} we need
not consider the cross term $-\langle S_gS_\lambda\rangle_{0>}$ which
does not provide any perturbative corrections to the action. In the
following, we shall only show how we compute the perturbative
corrections arising from the sine-Gordon perturbation
$-\frac{1}{2}\langle S_g^2\rangle_{0>}$, the unique term arising from
$-\frac{1}{2}\langle S_\lambda^2\rangle_{0>}$
\begin{eqnarray}
\Delta S_\lambda(\gamma T) = \int d{\bf r}dt \tilde{u}^2_<({\bf r},t) \Big[\frac{T\lambda^2\,d\ell}{8\pi K^3}\Big]
\end{eqnarray}
having been repeatedly derived in the 
literature\cite{Hwa,Nattermann-Tang,RostSpohn}.
Taking the Gaussian averages in equation (\ref{cumul}) leads to the following expression of the 
second cumulant $\Delta S_g[\tilde u,u]=-\frac{1}{2}\langle S_g^2\rangle_{0>}^c$,
\end{multicols}
\begin{eqnarray}
\Delta S_g[\tilde u^< u]  & = &  -\frac{1}{2}\,p^2g^2\int \!\!d{\bf r}dt\int\!\! d{\bf r}'dt'\;
\tilde u^<({\bf r},t)\tilde u^<({\bf r}',t')\tilde{\cal K}({\bf r}-{\bf r}',t-t')
\cos\big[p\big(u^<({\bf r},t) - u^<({\bf r}',t')\big) +\frac{pF}{\gamma}(t-t')\big]
\nonumber\\
& - &\frac{1}{2}\,p^3g^2\int \!\!d{\bf r}dt\int\!\! d{\bf r}'dt'
\;i\tilde u^<({\bf r},t) {\cal K}({\bf r}-{\bf r'},t-t')
\sin\big[p(u^<({\bf r},t)-u^<({\bf r}',t'))+\frac{pF}{\gamma}(t-t')\big] \; .
\label{dyncumul1}
\end{eqnarray} 
\begin{multicols}{2}
Here the dynamic kernels $\tilde{\cal K}({\bf r},t)$ and ${\cal K}({\bf r},t)$ are given by
\begin{eqnarray}
\tilde{\cal K}({\bf r},t) & = & \frac{1}{2}\big[1-\cosh\big(p^2 G_0^>({\bf r},t)\big)\big]
\nonumber\\
&-& \sinh\big(p^2 G_0^>({\bf r},t)\big) \; ,
\label{tildedynkernel}\\
{\cal K}({\bf r},t) & = & \mbox{e}^{-\frac{1}{2}\,p^2 C_0^>({\bf r},t)}R_0^>({\bf r},t) \; ,
\label{dynkernel}
\end{eqnarray}
where $R_0^>({\bf r},t)=\int_{{\bf q},\omega}^>\mbox{e}^{-i({\bf q}\cdot{\bf r}-\omega t)}/(i\gamma\omega+Kq^2)$ and
$C_0^>({\bf r},t)=\langle[u^>({\bf r},t)-u^>({\bf 0},0)]^2\rangle$ are the response and correlation functions,
respectively, and where the correlator $G_0^>({\bf r},t)=\langle u^>({\bf r},t)u^>({\bf 0},0)\rangle_{0>}$
is given by
\begin{eqnarray}
G_0^>({\bf r},t)=2{\gamma}T\int_{{\bf q},\omega}\frac{\cos[{\bf q}\cdot{\bf r}-\omega t]}{\gamma^2\omega^2 + K^2 q^4} \;.
\end{eqnarray}
We now decompose the sine and cosine in the integrand on the {\em rhs} of equation (\ref{dyncumul1}) according to
\end{multicols}
\begin{mathletters}
\begin{eqnarray}
\cos\big[p(u_<-u_<')+\frac{pF}{\gamma}(t-t')\big] & = &
\sin\big[p(u_< - u_<')\big]\cos\big[\frac{pF}{\gamma}(t-t')\big]+ 
\cos\big[p(u_<-u_<')\big]\sin\big[\frac{pF}{\gamma}(t-t')\big] \, ,
\\
\sin\big[p(u_<-u_<')+\frac{pF}{\gamma}(t-t')\big] & = &
\sin\big[p(u_<-u_<')\big]\cos\big[\frac{pF}{\gamma}(t-t')\big]+
\cos\big[p(u_<-u_<')\big]\sin\big[\frac{pF}{\gamma}(t-t')\big]\, .
\end{eqnarray}
\end{mathletters}
The kernels $\tilde{\cal K}({\bf r}-{\bf r'},t-t')$ and ${\cal K}({\bf r}-{\bf r'},t-t')$ being short ranged both in space and time, we
see that the major contribution to the action (\ref{dyncumul1}) comes from the regions ${\bf r}\simeq{\bf r'}$
and $t\simeq t'$ where $\big[u^<({\bf r},t)-u^<({\bf r}',t')\big]$ is small.
We therefore shall approximate
\begin{eqnarray}
\sin\big[p(u^<({\bf r},t)-u^<({\bf r}',t'))\big] & \simeq & p\,\big(u^<({\bf r},t)-u^<({\bf r}',t')\big) \; ,
\label{simplesine}\\
\cos\big[p(u^<({\bf r},t)-u^<({\bf r}',t'))\big] & \simeq & 1 - 
\frac{1}{2}\,p^2\,\big(u^<({\bf r},t)-u^<({\bf r}',t')\big)^2 \; ,
\label{simplecosine}
\end{eqnarray}
and
\begin{equation}
u^<({\bf r},t)-u^<({\bf r}',t') = (t-t')\,\partial_t u^< +({\bf r}-{\bf r'})_\alpha\,\partial_\alpha u^< 
+\frac{1}{2}({\bf r}-{\bf r'})_\alpha({\bf r}-{\bf r'})_\beta\partial_\alpha\partial_\beta u^< \; ,
\end{equation} 
upon which we obtain the following expression for the second cumulant $-\frac{1}{2}\langle S_1^2\rangle_{0>}^c$~:
\begin{eqnarray}
\Delta S[\tilde{u},u]= -\frac{1}{2}\langle S^2\rangle_{0>}^c[\tilde{u},u] = 
\Delta S(\gamma T) + \Delta S(\gamma) + \Delta S(K) + \Delta S(\lambda) + \Delta S(F) \; ,
\end{eqnarray}
where
\begin{mathletters}
\begin{eqnarray}
\Delta S(\gamma T) & = & \Delta S_\lambda(\gamma T) +
\frac{1}{2}p^2g^2\int d{\bf r}dt\;\tilde{u}^<({\bf r},t)\tilde{u}^<({\bf r},t)
\int d{\bf r}'dt'\;\tilde{\cal K}({\bf r}-{\bf r'},t-t')\cos\big[\frac{pF}{\gamma}(t-t')\big] \; ,
\label{delta_gammaT}\\
\Delta S(\gamma) & = & \frac{1}{2}\,p^4g^2\int d{\bf r}dt \; i\tilde u^<({\bf r},t)\,[\partial_t u^<({\bf r},t) ]
\int d{\bf r'}dt'\,(t-t')\,{\cal K}({\bf r}-{\bf r'},t-t')\cos\big[\frac{pF}{\gamma}(t-t')\big] \; ,
\label{delta_gamma}\\
\Delta S(K) & = & \frac{1}{4}\,p^4g^2\int d{\bf r}dt \; i\tilde u^<({\bf r},t)\,[-\nabla^2 u^<({\bf r},t)]
\int d{\bf r'}dt'\,({\bf r}-{\bf r'})^2\,{\cal K}({\bf r}-{\bf r'},t-t')
\cos\big[\frac{pF}{\gamma}(t-t')\big]  \; ,
\label{delta_K}\\
\Delta S(\lambda) & = & \frac{1}{4}\,p^5g^2\int d{\bf r}dt \; i\tilde u^<({\bf r},t)\,
\big[-\big(\nabla u^<({\bf r},t)\big)^2\big]
\int d{\bf r'}dt'\,({\bf r}-{\bf r'})^2\,{\cal K}({\bf r}-{\bf r'},t-t')
\sin\big[\frac{pF}{\gamma}(t-t')\big] \; ,
\label{delta_lambda}\\
\Delta S(F) & = & \frac{1}{2}\,p^3g^2\int d{\bf r}dt \; i\tilde u^<({\bf r},t)  
\int d{\bf r'}dt'\,{\cal K}({\bf r}-{\bf r'},t-t')\sin\big[\frac{pF}{\gamma}(t-t')\big]  \; .
\label{delta_F}
\end{eqnarray} 
\end{mathletters}
\begin{multicols}{2}
Here we pause a moment to indicate that if we use the complete expression of the kernel ${\cal K}({\bf r},t)$ 
\begin{eqnarray}
{\cal K}({\bf r},t) = \mbox{e}^{-\frac{1}{2}\,p^2 C_0^>({\bf r},t)}\;R_0^>({\bf r},t)
\end{eqnarray}
into equation (\ref{delta_F}) and let $b\to\infty$, then we obtain from equation (\ref{delta_F}) above 
the following expression for the friction force $F_{fr}$ due to the pinning potential to order $g^2$,
\begin{eqnarray}
F_{fr} &=& \frac{1}{2}\,p^3g^2
\int d{\bf r'}dt'\, \mbox{e}^{-\frac{1}{2}\,p^2 C_0({\bf r}-{\bf r}',t-t')} \times\nonumber\\
&\times&\;R_0({\bf r}-{\bf r'},t-t')\sin\big[\frac{pF}{\gamma}(t-t')\big] \; ,
\end{eqnarray}
which leads directly to the perturbative result (\ref{vfcharPT}) of the text.

We now go back to our dynamic RG recursion relations, 
(\ref{delta_gammaT})-(\ref{delta_F}).
In the dynamic kernels of equations (\ref{tildedynkernel})-(\ref{dynkernel}), 
we expand
\begin{mathletters}
\begin{eqnarray}
\tilde{\cal K}({\bf r},t) & = & - p^2G_0^>({\bf r},t) 
- \frac{1}{4}p^4\big(G_0^>({\bf r},t)\big)^2 \; ,
\label{simplif-tildeK}
\\
{\cal K}({\bf r},t) & = & R_0^>({\bf r},t) -\frac{1}{2}\, p^2C_0^>({\bf r},t)R_0^>({\bf r},t) \; ,
\label{simplif-K}
\end{eqnarray}
\end{mathletters}
and keep only the second term on the {\em rhs} of the above equations
(the first term gives a vanishing contribution, for reasons which are
identical to those explained after equation (\ref{vanishes}) of
appendix A).  Now, from equations (\ref{delta_gamma})-(\ref{delta_F}),
we see that the perturbative corrections to the bare parameters of the
theory are given by the flows
\begin{eqnarray}
\frac{d(\gamma T)}{d\ell}\big|_{pert} & = & 
\frac{T\lambda^2}{8\pi K^3} +
\,p^2g^2 \int d{\bf r}\,dt \; \tilde{\cal K}({\bf r},t)\cos\big(\frac{pF}{\gamma}\,t\big) \, ,
\nonumber\\
\frac{d\gamma}{d\ell}\big|_{pert} & = & 
\frac{1}{2}\,p^4g^2 \int d{\bf r}\,dt \; t{\cal K}({\bf r},t)\cos\big(\frac{pF}{\gamma}\,t\big) \; , 
\nonumber\\
\frac{dK}{d\ell}\big|_{pert} & = & \frac{1}{4}\,p^4g^2 \int d{\bf r}\,dt \; r^2{\cal K}({\bf r},t)
\cos\big(\frac{pF}{\gamma}\,t\big) \; ,
\nonumber\\
\frac{d\lambda}{d\ell}\big|_{pert} & = & \frac{1}{4}\,p^5g^2 \int d{\bf r}\,dt \; r^2{\cal K}({\bf r},t)
\sin\big(\frac{pF}{\gamma}\,t\big) \; ,
\nonumber\\
\frac{dF}{d\ell}\big|_{pert} & = & \frac{1}{2}\,p^3g^2 \int d{\bf r}\,dt \; {\cal K}({\bf r},t)
\sin\big(\frac{pF}{\gamma}\,t\big) \; .
\nonumber
\end{eqnarray}
Using equations (\ref{simplif-tildeK})-(\ref{simplif-K}),
the above recursion relations become
\begin{mathletters}
\begin{eqnarray}
\frac{d}{d\ell}(\gamma T) & = & \big[\frac{T\lambda^2}{8\pi K^3}\! +
\!\frac{Tp^6g^2}{16\pi K^3\Lambda^4}\frac{1}{1+f^2}\big](\gamma T) \; ,
\\
\frac{d\gamma}{d\ell} & = & \frac{Tp^6g^2}{16\pi K^3\Lambda^4}\frac{1-f^2}{(1+f^2)^2}\,\gamma \; ,
\\
\frac{dg}{d\ell} & = & \big(2-\frac{Tp^2}{4\pi K}\big)\,g   \; ,
\\
\frac{dK}{d\ell} & = & \frac{Tp^6g^2}{16\pi K^2\Lambda^4}\,\frac{2-3f^2-f^4}{(1+f^2)^3} \; ,
\\
\frac{d\lambda}{d\ell} & = & \frac{Tp^7g^2}{16\pi K^2\Lambda^4}\,\frac{f(f^2+5)}{(1+f^2)^3} \; ,
\\
\frac{dF}{d\ell} & = & \frac{\lambda T\Lambda^2}{4\pi K}-\frac{Tp^5g^2}{8\pi K^2\Lambda^2}\frac{f}{1+f^2} \; .
\end{eqnarray}
On the other hand, we know from equations (\ref{space_rescaling})-(\ref{field_rescaling}) that the rescaling of 
fields and space and time variables produces the recursion relations
\begin{eqnarray}
\frac{d(\gamma T)}{d\ell}\big|_{resc} & = & \frac{d\gamma}{d\ell}\big|_{resc} = \frac{dK}{d\ell}\big|_{resc} =
\frac{d\lambda}{d\ell}\big|_{resc} = 0 \; ,
\nonumber\\
\frac{dF}{d\ell}\big|_{resc} & = & 2F \; .
\end{eqnarray}
\end{mathletters}
Using the recursion relations above along with the fact that, in a renormalization group transformation,
\begin{eqnarray}
\frac{d}{d\ell} = \frac{d}{d\ell}\Big|_{pert} + \frac{d}{d\ell}\Big|_{resc} \; ,
\end{eqnarray}
leads directly to equations (\ref{recgamma})-(\ref{recF}) of the text.

\end{multicols}

\begin{references}
\bibitem{Remark1} The situation here is quite analogous to the
  lack of qualitative distinction and therefore to the rounding of the
  phase transition between a paramagnetic and ferromagnetic phases in
  the presence of a finite external magnetic field $H$, with
  magnetization $M(H)$ finite in both phases. Nevertheless, in the
  limit of a vanishing field, the equation of state $M(H,T)$ displays
  a qualitatively distinct behavior as a function of $H$, exhibiting a
  diverging differential susceptibility for $T<T_c$ and a finite one
  for $T>T_c$.
\end{references}
\end{document}